\documentclass{aa} 

\usepackage{graphicx}
\usepackage{txfonts}
\usepackage{xcolor}
\usepackage{hyperref}
\usepackage{tablefootnote}
\usepackage{orcidlink}
\usepackage{academicons}
\usepackage[normalem]{ulem}
\begin{document}

   \title{JWST MIRI/MRS in-flight absolute flux calibration and tailored fringe correction for unresolved sources}

   \author{Danny Gasman\inst{1}\orcidlink{0000-0002-1257-7742}
          \and
          Ioannis Argyriou\inst{1}\orcidlink{0000-0003-2820-1077}
          \and
          G. C. Sloan\inst{2,3}\orcidlink{0000-0003-4520-1044}
          \and
          Bernhard Aringer\inst{4}\orcidlink{0000-0001-9848-5410}
          \and
          Javier Álvarez-Márquez\inst{5}\orcidlink{0000-0002-7093-1877}
          \and
          Ori Fox\inst{2}\orcidlink{0000-0003-2238-1572}
          \and
          Alistair Glasse\inst{6}\orcidlink{0000-0002-2041-2462}
          \and
          Adrian Glauser\inst{7}\orcidlink{0000-0001-9250-1547}
          \and
          Olivia C. Jones\inst{6}\orcidlink{0000-0003-4870-5547}
          \and
          Kay Justtanont\inst{8}\orcidlink{0000-0003-1689-9201}
          \and
          Patrick~J.~Kavanagh\inst{9}\orcidlink{0000-0001-6872-2358}
          \and
          Pamela Klaassen\inst{6}\orcidlink{0000-0001-9443-0463}
          \and
          Alvaro Labiano\inst{5,10}\orcidlink{0000-0002-0690-8824}
          \and
          Kirsten Larson\inst{2}\orcidlink{0000-0003-3917-6460}
          \and
          David~R.~Law\inst{2}\orcidlink{0000-0002-9402-186X}
          \and
          Michael Mueller\inst{11}\orcidlink{0000-0003-3217-5385}
          \and
          Omnarayani Nayak\inst{2}\orcidlink{0000-0001-6576-6339}
          \and
          Alberto Noriega-Crespo\inst{2}\orcidlink{0000-0002-6296-8960}
          \and
          Polychronis Patapis\inst{7}\orcidlink{0000-0001-8718-3732}
          \and
          Pierre Royer\inst{1}\orcidlink{0000-0001-9341-2546}
          \and
          Bart Vandenbussche\inst{1}\orcidlink{0000-0002-1368-3109}
          }

   \institute{Institute of Astronomy, KU Leuven, Celestijnenlaan 200D, 3001 Leuven, Belgium
   \and
   Space Telescope Science Institute, 3700 San Martin Drive, Baltimore, MD 21218, USA.
   \and
   Department of Physics and Astronomy, Univ. of North Carolina at Chapel Hill, Chapel Hill, NC 27599-3255, USA
   \and
   Department of Astrophysics, Univ. of Vienna, Türkenschanzstraße 17, A-1180 Wien, Austria
   \and
   Centro de Astrobiolog\'ia (CAB), CSIC-INTA, ESAC, Carretera de Ajalvir km4, 28850 Torrej\'on de Ardoz, Madrid, Spain
   \and
   UK Astronomy Technology Centre, ROE, Blackford Hill, Edinburgh, EH9 3HJ, UK
   \and 
   Institute for Particle Physics and Astrophysics, ETH Zurich, Wolfgang-Pauli-Str 27, 8093 Zurich, Switzerland
   \and
   Dept. of Space, Earth and Environment, Chalmers University of Technology, Onsala Space Observatory, S-43992 Onsala, Sweden
   \and
   School of Cosmic Physics, Dublin Institute for Advanced Studies, 31 Fitzwilliam Place, Dublin 2, Ireland
   \and
   Telespazio UK for the European Space Agency, ESAC, Camino Bajo del Castillo s/n, 28692 Villanueva de la Ca\~nada, Spain
   \and
   Kapteyn Astronomical Institute, Rijksuniversiteit Groningen, Postbus 800, 9700AV Groningen, Netherlands
   \\
   \email{danny.gasman@kuleuven.be}
    }

   \date{Received December 7, 2022; accepted March 2 2023}

 
  \abstract
   {The Medium Resolution Spectrometer (MRS) is one of the four observing modes of JWST/MIRI. Using JWST in-flight data of unresolved (point) sources, we can derive the MRS absolute spectral response function (ASRF) starting from raw data. Spectral fringing, caused by coherent reflections inside the detector arrays, plays a critical role in the derivation and interpretation of the MRS ASRF. The fringe corrections implemented in the current pipeline are not optimal for non-extended sources, and a high density of molecular features particularly inhibits an accurate correction.}
   {In this paper we present an alternative way to calibrate the MIRI/MRS data. Firstly, we derive a fringe correction that accounts for the dependence of the fringe properties on the MIRI/MRS pupil illumination and detector pixel sampling of the point spread function. Secondly, we derive the MRS ASRF using an absolute flux calibrator observed across the full 5 to 28 $\mu m$ wavelength range of the MRS. Thirdly, we apply the new ASRF to the spectrum of a G dwarf and compare it with the output of the JWST/MIRI default data reduction pipeline. Finally, we examine the impact of the different fringe corrections on the detectability of molecular features in the G dwarf and K giant.}
   {The absolute flux calibrator HD~163466 (A-star) was used to derive tailored point source fringe flats at each of the default dither locations of the MRS. The fringe-corrected point source integrated spectrum of HD~163466 was used to derive the MRS ASRF using a theoretical model for the stellar continuum. A cross-correlation was run to quantify the uncertainty on the detection of CO, SiO, and OH in the K giant and CO in the G dwarf for different fringe corrections.}
   {The point-source-tailored fringe correction and ASRF are found to perform at the same level as the current corrections, beating down the fringe contrast to the sub-percent level in the G dwarf in the longer wavelengths, whilst mitigating the alteration of real molecular features. The same tailored solutions can be applied to other MRS unresolved targets. Target acquisition is required to ensure the pointing is accurate enough to apply this method. A pointing repeatability issue in the MRS limits the effectiveness of the tailored fringe flats is at short wavelengths. Finally, resulting spectra require no scaling to make the sub-bands match, and a dichroic spectral leak at 12.2 micron is removed.}
   {}

    
    \titlerunning{JWST MIRI/MRS in-flight absolute flux calibration for unresolved sources}
    \authorrunning{D. Gasman}
   \maketitle
%

\section{Introduction}
\label{sec:introduction}

During the six-month commissioning phase after the launch of the James Webb Space Telescope \citep[\textit{JWST},][]{ref:06GaMaCl,ref:22jwst}, the Medium Resolution integral field Spectrometer \citep[MRS,][]{ref:15WePeGl} of the Mid-Infrared Instrument \citep[MIRI,][]{ref:23WrRiGl} observed a variety of targets for the purpose of verifying its scientific performance (Argyriou et al., in prep.). The commissioning targets included spatially unresolved (point) sources, namely stars with different effective temperatures, as well as spatially resolved sources, such as galaxies and planetary nebulae; and it has since been the topic of a variety of scientific works \citep[e.g.][]{ref:corinos,ref:22GaRiAl,ref:22MiBiPa,ref:22AlLaGu,ref:22Lau}.

The MRS is an integral field spectrometer, meaning it provides both spatial and spectral information. The 5--28~$\mu m$ range of the MRS is divided into four channels (labelled 1--4 with increasing wavelength), each divided into three sub-bands (labelled A--C or SHORT--LONG with increasing wavelength). After reducing the data separately per sub-band, this results in 12 separate spectral `cubes', with two spatial axes and one spectral axis (see Law et al., in prep.).

In the spectral direction, the MIRI MRS shows high amplitude fringes caused by Fabry-Pérot interference between the reflective layers of the detectors, the depth of which can cause errors up to 30\% of the total flux if left uncorrected \citep{ref:20ArRiRe,ref:20ArWeGl}. Fringing has been a systemic issue for instruments on other space telescopes, such as the Space Telescope Imaging Spectrograph on board the \textit{Hubble Space Telescope} \citep{ref:03MaHiGu}, the Short Wavelength Spectrometer on board the \textit{Infrared Space Observatory} \citep{ref:03KeBeLu}, and the InfraRed Spectrograph (IRS) on board \textit{Spitzer} \citep{ref:00LaDi,ref:03LaBo}. The MRS spectral resolution of 4000-1500 over the 5 to 28~$\mu m$ wavelength range \citep{ref:21LaArAl}, together with the $\sim$500~$\mu m$ thick detectors result in the fringe frequencies being spectrally resolved.

When observing unresolved (point) sources, the MIRI pupil is non-uniformly illuminated. This causes a significant change in the fringe depth and phase as a function of the part of the point spread function (PSF) that is sampled by the detector pixels \citep{ref:20ArWeGl}. For spatially extended sources, the MIRI pupil is uniformly illuminated and the fringe properties average out, varying mainly with wavelength. The effectiveness of applying a 2D static fringe flat, derived from a spatially extended source (uniform pupil illumination), to correct the fringes of an unresolved point source (non-uniform pupil illumination) is limited by the underlying physics in question, and can substantially change the pristine shape of the fringes and hence the shape of real spectral lines.

The \textit{JWST} data reduction pipeline\footnote{\url{https://jwst-pipeline.readthedocs.io/en/latest/}} contains two methods to remove fringes from MRS spectra \citep[a first description of the pipeline can be found in][]{ref:16LaAzBa}: a fringe flat derived from spatially extended sources (Mueller et al., in prep) and a residual fringe correction that iteratively finds and removes remaining periodic features in the spectrum that match the expected fringe frequencies based on the geometric and refractive properties of the detectors (Kavanagh et al., in prep). Spatially extended sources in the context of MRS calibration included (1) an 800K black-body source used during the MIRI Flight Model (FM) test campaign, (2) the MRS internal calibration source (hot tungsten wire filament with a grey-body temperature of 800 K), and  (3) the Cat's Eye nebula NGC6543 observed during the JWST commissioning phase. Due to the beam from the internal calibration source being significantly more collimated than the 800K black-body source placed external to MIRI, resulting in a larger fringe amplitude and detectable phase offsets, the data from the internal calibration source could not be used to derive fringe flats. For this reason a combination of fringe flats based on the 800K black-body source and the Cat's Eye nebula are used for the in-flight calibration reference files, as is  discussed later in this paper. Taken together, the fringe flats based on spatially extended sources and the residual fringe correction have been shown to reduce fringe contrast to below 6\% across the MRS wavelength range (Kavanagh et al., in prep.). However, a significant portion of MIRI science programmes require an uncertainty on the continuum below 1\% (S/N of 300 for a 3$\sigma$ detection). Due to the empirical nature of the residual fringe correction correction, it cannot recover the intrinsic detector absolute spectral response function (ASRF).

There are three overarching issues remaining after commissioning relevant to the work presented here. Firstly, for sources with dense molecular bands, either the bands may be identified as fringes by the residual fringe correction algorithm, or they may cause the algorithm to fail to identify fringes; resulting in either the removal of real features in the process or the failure to remove fringes, respectively. As such, we seek to derive a tailored point source fringe flat (PSFF; this acronym was chosen to impress the link between the instrumental PSF and the fringes) from a spectrally `boring' source: a predominantly featureless flux calibration standard placed in the exact same position on the detector as a science target of interest.

Secondly, the in-flight MRS spectrophotometric response solution contains all the noise and systematics from the flight model (FM) test campaign. Due to various problems with the test set-up, for example imperfections in the extended fringe flats used and the presence of a water feature (for more details see \citealt{phdthesisYannis}, and Sect. \ref{subsec:photom} in this paper), systematics were introduced into the spectrophotometric calibration files. During commissioning, the MRS ASRF was derived by applying a 1D correction to the 2D spectrophotometric response on the detector image plane derived during the FM test campaign. The 1D correction was derived by extracting the point source integrated spectrum of HD~163466 using the ground RSRF, estimating the spectral continuum, and dividing the spectral flux density of the estimated continuum by the values expected based on a theoretical model of the continuum of HD~163466 \citep{ref:17BoMeFl,ref:20BoHuRa}. As a result, some of the systematic uncertainties linked to the ground RSRF are still present in the MRS reference files used to calibrate MRS in-flight data. In this paper we show how to derive a point source optimised MRS ASRF from raw data.

Thirdly, the pointing of the instrument, even with target acquisition, has an accuracy of $\sim$30~mas. Due to changes in the sampling of the PSF with changes in pointing, the depth and phase of the fringes will change. We show that, ultimately, this will be the limiting factor for the effectiveness of the PSFF corrections derived here.

With the new PSFF and ASRF solutions at hand, we  used the MRS spectra of HD~37122 and HD~159222 as our two test cases. The K giant star HD~37122 was observed during commissioning as part of the target acquisition test; and the solar analogue G dwarf HD~159222 was one of the targets used to examine the MRS spectrophotometric sensitivity and stability, alongside the previously mentioned A star HD~163466 \citep{ref:22GoBoSl}.

To quantify how well the different methods perform, and to discuss the implications on the science, we performed cross-correlations to compare the K giant spectra to the synthetic molecular spectra of CO, SiO, and OH, all of which are expected to be detectable \citep{ref:03DeVaWa,ref:15SlHeCh}; and CO with the G dwarf, which should be present in a star of this spectral type \citep{ref:02HeShPr,ref:10ArDyMa}.

In Sect. \ref{sec:methods} we describe the MRS in-flight observations used in this paper, how the PSFF solutions are derived, and the different methods of reducing the in-flight data. In addition, we show how the MRS ASRF solution is derived, including the correction of a spectral leak feature at 12.2~$\mu m$. In Sect. \ref{sec:results} we show the fringe-and-ASRF calibrated spectra of the K giant and G dwarf, along with the implications on the cross-correlation results. The effects on the science of sources rich in molecular bands is discussed in Sect. \ref{sec:discussion}. Finally, we present our conclusions in Sect. \ref{sec:conclusions}.

\section{Methods}
\label{sec:methods}

\subsection{Data used}
\label{subsec:data}
Table \ref{tab:prop} gives the information for each of the observing programs carried out during commissioning for the targets used in this paper. This includes the program identification numbers (PIDs) from the Astronomer Proposal Tool (APT)\footnote{\url{https://www.stsci.edu/scientific-community/software/astronomers-proposal-tool-apt}}.

The two most suitable publicly available calibration stars for the analysis in this paper observed during commissioning are HD~37122 and HD~159222. HD~37122 is a K2III C star located in the Large Magellanic Cloud (LMC) and observed as part of the Surveying the Agents of a Galaxy’s Evolution (SAGE) survey \citep{ref:06MeGoIn}. K stars have been used as spectrophotometric calibrators in the past, for example for \textit{Spitzer} \citep{ref:04HoRoCl} by comparing them to spectral templates \citep{ref:03CoMeHa} and synthetic spectra \citep{ref:04DeMoAp}. However, \citet{ref:15SlHeCh} suggest that our current inability to predict the strength of molecular bands in K giants makes them less suited for calibration purposes due to the wealth of molecular bands including CO, SiO, and OH. Instead, HD~37122 was used to test the target acquisition of MIRI MRS, due to its brightness in the mid-infrared. HD~159222 has been considered a representative solar analogue, in terms of age and spectrophotometric properties \citep{ref:04SoTr,ref:16MaBlCh}; however, \citet{ref:14MeSiSi} find that it has a higher temperature and higher metal abundances, and is overall more evolved than the Sun. Due to the lack of strong features at the longer wavelengths, it allows   a direct comparison of the different fringe removal methods, without convoluted results due to molecular features. It was observed twice, with a week between observations (indicated by `repeat visit' in Table~\ref{tab:prop}), in order to test the spectrophotometric repeatability.

For the derivation of the PSFF reference files, a relatively featureless star is used, the A star HD~163466, which was observed to derive the spectrophotometric correction factors during commissioning \citep{ref:22GoBoSl}. Additionally, it was one of the standard stars used for the calibration of \textit{Spitzer} \citep{ref:05ReMeCo,ref:07EnBlSu,ref:15SlHeCh}. A single dedicated background observation associated with PID~1050 was performed to use with HD~163466.

\begin{table}
\centering
\caption{Properties and programme information of the three stars examined in this work. \textit{1}: \citet{ref:16gaia,ref:21gaia,ref:21gaia2}; \textit{2}: \citet{ref:03GrCoGa}. Groups per integration show sub-band and/or channel when a different set-up was used for different bands.}
\label{tab:prop}
\resizebox{\columnwidth}{!}{%
\begin{tabular}{llll}
ID            & HD 37122$^{1}$                   & HD 159222$^{2}$ & HD 163466$^{2}$                   \\ \hline \hline
RA        & 05 30 00.7691 & 17 32 00.9923 & 17 52 25.3741
\\ 
DEC     & -69 58 31.871  & +34 16 16.131 & +60 23 46.94
\\
Spectral type & K2III C                    & G1V      & A7Vm                   
\\
$T_{\textrm{eff}}$ [K]     &   $\sim$4000    & 5790 & $\sim$9000
\\
K mag             & 5.13                       & 5.05  & 6.34
\\ \hline \hline
APT PID                                                                         & \href{https://www.stsci.edu/jwst/science-execution/program-information.html?id=1029}{1029}                                                                                & \href{https://www.stsci.edu/jwst/science-execution/program-information.html?id=1050}{1050}  & \href{https://www.stsci.edu/jwst/science-execution/program-information.html?id=1050}{1050}                                                                            \\
PI                                                                          & A. Glasse                                                                           & B. Vandenbussche   &    "                                                            \\ \hline
Date                                                                        & May 22-23, 2022                                                                  & \begin{tabular}[c]{@{}l@{}}June 6, 2022 \\ (first visit)\\ June 13, 2022 \\ (repeat visit)\end{tabular} & June 8-9, 2022 \\ \hline
Dithers\tablefootnote{\url{https://jwst-docs.stsci.edu/jwst-mid-infrared-instrument/miri-operations/miri-dithering/miri-mrs-dithering}}                                                                     & \begin{tabular}[c]{@{}l@{}}4 (positive)\\ 4 (negative)\end{tabular}                 & 4 (negative) & \begin{tabular}[c]{@{}l@{}}4 (positive)\\ 4 (negative)\end{tabular}    \\ \hline
\begin{tabular}[c]{@{}l@{}}Groups/integrations\\ (per sub-band)\end{tabular} & \begin{tabular}[c]{@{}l@{}}12/7 (1/2 A, B)\\ 15/6 (1/2 C)\\ 30/3 (3/4)\end{tabular} & \begin{tabular}[c]{@{}l@{}}15/6 (1/2)\\ 45/2 (3/4)\end{tabular}    & 50/2                            \\ \hline
Time per dither {[}s{]}                                                     & \begin{tabular}[c]{@{}l@{}}249.754\\ 263.629\\ 255.304\end{tabular}                 & \begin{tabular}[c]{@{}l@{}}263.629\\ 252.529\end{tabular} & 280.279 \\ \hline \hline
& & & 1050 background \\
Dithers & & & 2 (negative) \\
Groups/integrations & & & 50/2 \\
Time per dither {[}s{]} & & & 560.558
\end{tabular}%
}
\end{table}

The synthetic spectra showing the transitions of CO, SiO, and OH used in this paper are based on hydrostatic models taken from the COMARCS grid for cool stars \citep{ref:16ArGiNo}. In the case of HD~37122 we used a computation with $\rm T_{eff} = 4000~K$ and $\rm log(g~[cm/s^2]) = 1.54$. For HD~159222 the corresponding values are $\rm T_{eff} = 5800~K$ and $\rm log(g~[cm/s^2]) = 4.40$. Both models have solar abundances and one solar mass. The microturbulent velocity was set to 2.5~km/s. We   note that these properties were chosen to obtain spectra similar to those of the two selected stars. No fit of the MRS or any other observed data has been made. The molecular spectra were calculated from the original model structures by including only the line opacities of the
corresponding species in the radiative transfer. For CO the data are   from \citet{ref:15LiGoRo}, for SiO from ExoMol \citep{ref:13BaYuTe}, and for OH from HITEMP \citep{ref:10RoGoBa}. The original resolution of R~=~200000 was reduced by a convolution to match the MRS observations.

\subsection{Fringe flats and residual fringe correction}
The extended fringe flats contain the normalised fringe pattern for each MRS spectral band across the detector image plane. The \textit{JWST} pipeline divides the signal on the detector image plane by the extended fringe flats. Above 10~$\mu m$, the extended fringe flats were derived based on observations of the Cat's eye nebula (NGC~6543, APT PID \href{https://www.stsci.edu/jwst/science-execution/program-information.html?id=1047}{1047}). The spatial structure of the nebula does not impact the derivation of the fringe flat as long as there is useful signal; the fringe properties themselves do not depend on the intensity of the source. The presence of spectral features does affect the definition of the spectral baseline, and thus the fringe properties. Spectral lines were masked out where possible (i.e. not included in the fitting), however, the trade-off is that fewer samples are available to use in the derivation of the fringe flat, increasing the sinusoidal model fit uncertainty; the detailed information can be found in Mueller et al. (in prep). At shorter wavelengths where the S/N of NGC~6543 was insufficient, no update to the extended fringe flats was derived during commissioning, and hence they are based on the FM ground test campaign solutions. Due to the beating of two frequencies produced in different layers of the MRS detectors \citep{ref:20ArRiRe} the fringe flats are least accurate in the 10 to 13~$\mu m$ region. In addition, a third fringe component occurs in the MRS dichroic filters, evident as a very low amplitude high frequency fringes above 11.5~$\mu m$, which the extended fringe flat does not incorporate.

The extended fringe flat correction reduces the contrast of the fringes, but residuals will remain as described in Sect. \ref{sec:introduction}. The \texttt{residual\_fringe} method in the pipeline uses an empirical sine model fitting method to further reduce the contrast of the fringing, under the assumption that the residual fringes are small in amplitude following correction by the extended fringe flat. To prevent the removal of spectral features in the fitting process, an automated spectral feature finding algorithm is used. The algorithm computes a periodogram from a spectrum and looks for prominent frequencies to remove in a range based on the understanding of the geometric and refractive properties of the MRS detectors. As in \citet{ref:03KeBeLu}, a Bayesian evidence loop is then used to identify the optimum number of sinusoids to include in the fit in order to prevent  overfitting (Kavanagh et al., in prep.). This algorithm can be applied directly after the \textit{JWST} pipeline \texttt{fringe} step, in 2D, and/or after spectral extraction, in 1D. For this work we apply both the 2D correction at the end of the \texttt{calspec2} module of the pipeline, and the post-pipeline residual fringe correction (i.e. on the 1D extracted spectrum) in order to be consistent with how the vast majority of MIRI science teams are reducing the MRS data.

\subsection{PSFF: Fringe flat tailored to unresolved (point) sources}
\label{subsec:ps_fringe_flat}

\begin{figure}[h]
    \centering
    \includegraphics[width=\columnwidth]{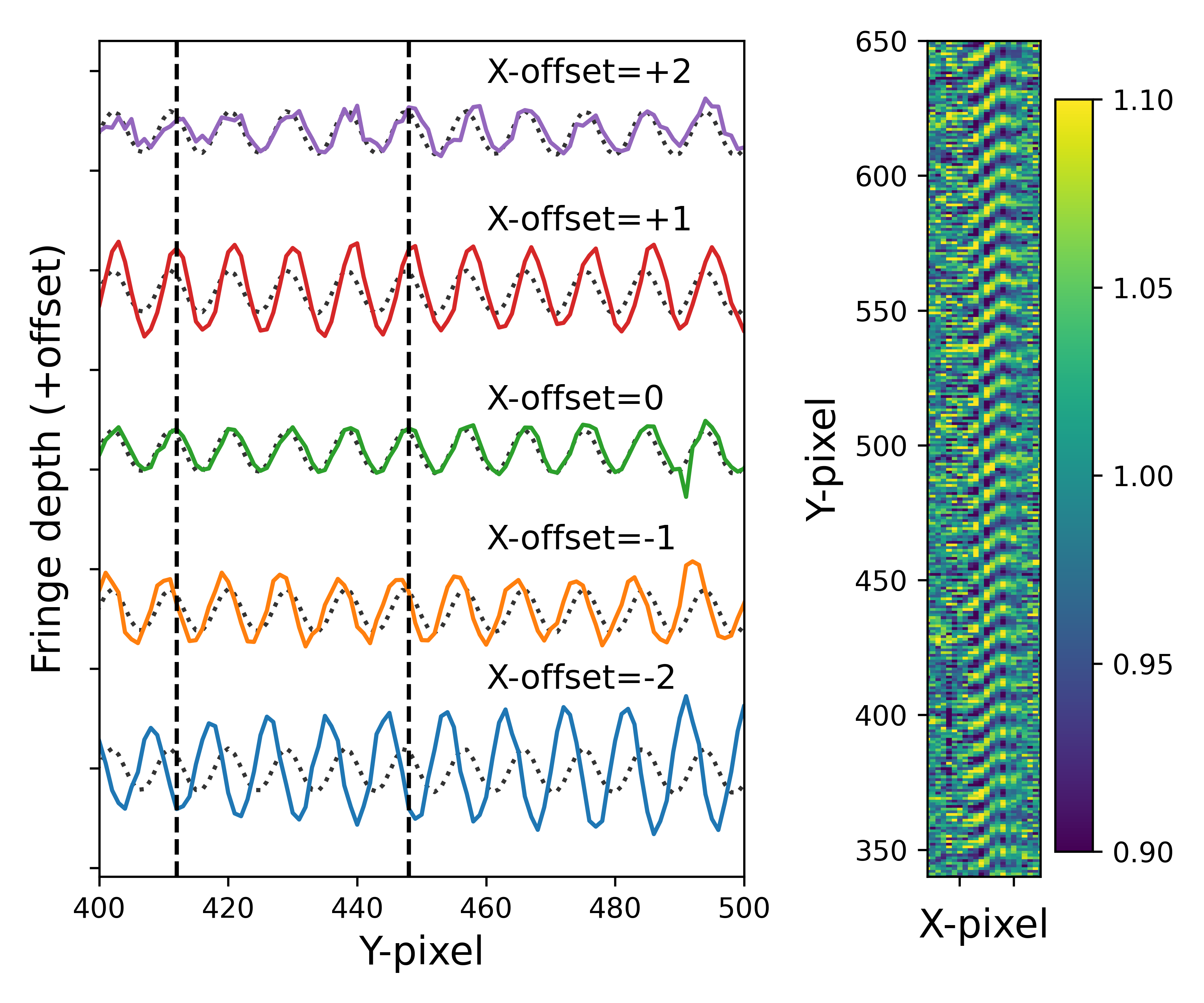}
    \caption{Change in fringe pattern with X-pixel offset from PSF peak (left), and selected area (X-pixel range of 314--335, Y-pixel range of 340--650) of the PSFF fringe flat detector image (right). The data in the left panel is offset (with respect to each other) in the y-direction by 0.4, and the black vertical lines are the location of two fringe peaks at X-offset=1. The extended fringe flat is plotted as a grey dotted line. The change in phase is visible in both the left and right panels, while the change in depth is most visible on the left. The MRS's IFU uses an image slicer to separate the different parts of the field of view. In this figure, only a single slice is shown.}
    \label{fig:flat}
\end{figure}

The source used for the purpose of deriving the PSFF reference files is the A star HD~163466. To produce a PSFF for the unresolved source HD~163466, spectral features were removed from the data using \texttt{sigma\_clip} in \textit{Astropy} \citep{astropy:2013,astropy:2018} with $\sigma~=~3$, after which the continuum was removed per detector column using a running mean. Any clipped sections are set to a value of 1. The resulting fringe pattern in the spectral direction, as a function of pixel-offset with respect to the PSF peak, is shown in the left panel of Fig.~\ref{fig:flat}. As a comparison, the extended fringe flat from commissioning is shown as a grey dotted line. A section of the resulting 2D PSFF fringe flat is shown in the right panel of Fig.~\ref{fig:flat}, where the point source pattern is visible as a streak of zebra stripes.

We have already discussed that the fringe pattern of an unresolved source depends on how the incoming wavefront from the MIRI pupil, on the detector, is sampled by the detector pixels. The quality of an empirically derived fringe flat for an unresolved source is limited by our ability to reproduce the sub-pixel pointing between calibrator (e.g. HD~163466) and the science target of interest. The left panel in Fig.~\ref{fig:flat} shows a part of the PSFF fringe flat at the centre of the PSF for different detector columns (offsets in X-pixel direction, PSF centre with offset=0), and demonstrates that  the depth and phase both change depending on the part of the PSF that is sampled. In the right panel in the 2D PSFF fringe flat, this phase shift is clearly visible as a diagonal zebra pattern. In the case where the MIRI pupil is illuminated uniformly (extended source illumination) the zebra stripes are approximately horizontal \citep{ref:20ArWeGl}.

As shown in the right panel of Fig.~\ref{fig:flat}, the part of the detector observing background is flat and noisy in the derived PSFF. In parallel, the parts of the detector on which the source is dispersed show the zebra pattern corresponding to fringe peaks and troughs. These peaks and troughs, due to dividing by the continuum, are now expressed in fringe depth as a fraction of the assumed continuum.

In this work, the PSFF is only applied to the G dwarf, and not the K giant. The K giant data was taken earlier in commissioning, prior to an update to the instrument's astrometric calibration. This causes the pointing to be inconsistent with the current pattern, and the offset to be larger than a pixel. However, due to the presence of dense molecular bands, it is a perfect test case for the removal of molecular features by the residual fringe correction.

\subsection{Spectral extraction}
\label{subsec:1d_extract}
\begin{figure}[h!]
    \centering
    \includegraphics[width=0.8\columnwidth]{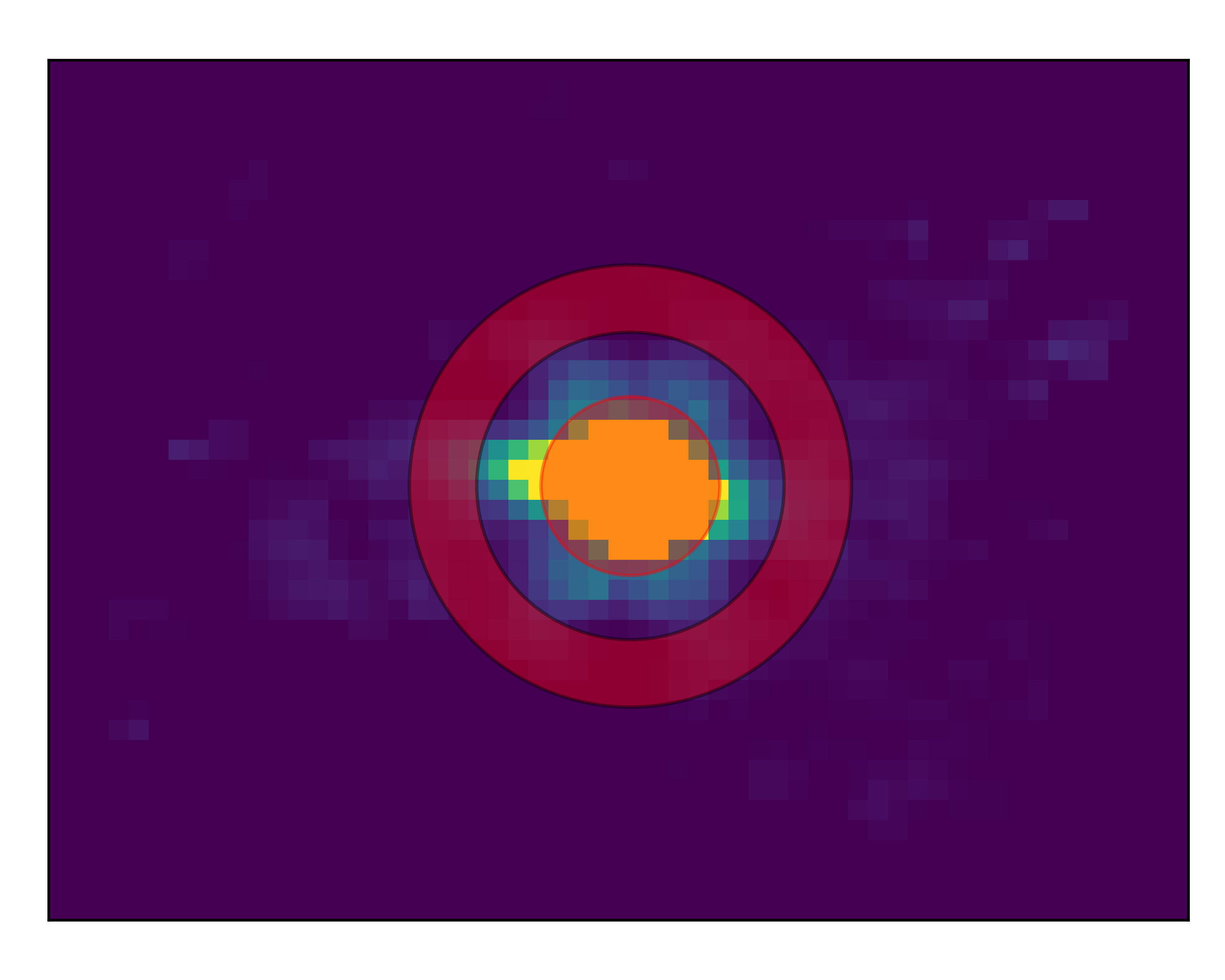}
    \caption{Example showing a cube image of band 1A image with a circular aperture and annulus overplotted.}
    \label{fig:extract_example}
\end{figure}

The data from each observing programme were reduced using the \textit{JWST} calibration pipeline v1.8.3;  the outlier rejection in step 3 was skipped because this step has been found to give spurious results due to the spatially undersampled PSF. One-dimensional spectra were extracted from the MRS 3D spectral cubes created using a 3D drizzling algorithm (Law et al. in prep.). The default 1D extraction of the \textit{JWST} pipeline was skipped. Instead, we extracted the spectra from the cubes manually, using a growing circular aperture centred on the PSF of 2.5$\lambda/D$ at each wavelength (radii from 0.470--2.757 arcsec). Due to instances of the centroid of the PSF being offset from the location expected by the \textit{JWST} pipeline, the aperture can be placed off-source. Skipping the step ensures that this is not an issue. A growing circular annulus was defined to subtract the background. The photometric extraction is illustrated in Fig. \ref{fig:extract_example}. To compute the 1D integrated point source spectrum, the signal within the aperture is summed, and the median signal in the annulus, scaled by the aperture area, is subtracted. An aperture correction factor is applied to account for the flux not included in the aperture and for the fraction of the PSF present inside the annulus. These values are consistent with those presented in Argyriou et al. (in prep).

\subsection{Spectrophotometric calibration and the PSFF}
\label{subsec:photom}
Before we can discuss the performance of the different fringe corrections, we must first discuss the spectrophotometric calibration solution as it is influenced by the fringe flat applied.

\subsubsection{MRS ground-based response solution systematics}
As we mention above,  the commissioning spectrophotometric calibration files are plagued by systematics introduced by ground testing. In \citet{phdthesisYannis} the systematic uncertainties linked to the ground RSRF solutions are discussed. We provide a summary here, along with the corresponding section in \citet{phdthesisYannis} in brackets. The MRS RSRF was derived using a 800~K black-body source that produced a spatially homogeneous extended illumination pattern on the detectors. Based on these observations, a set of 2D static extended fringe flats was derived and these flats were used as the reference fringe correction for all subsequent datasets collected over the multiple JWST/MIRI test campaigns that followed (Sect. 6.2 and 7.1.3). In order to derive the MRS RSRF, assumptions were made about the spectral flux density of the calibration source (starting from Planck's law), the geometry of the illumination by the MIRI Telescope Simulator (MTS) reaching the MIRI pupil, and the reflectivity curves of the MTS surfaces (Sect. 7.1.1). A water feature was present at 8~$\mu m$, which was folded into the MRS RSRF (Sect. 7.1.3). Electronic ghosts introduced part of the flux measured in detector column $c$ to detector column $c+4$ (Sect. 2.4.2).

In order to extract a point source spectrum with the correct values of spectral flux density (in units of jansky), a precise and accurate spectrophotometric response solution (hereafter PHOTOM) is required. As mentioned previously, the ground-based solution used to derive the commissioning PHOTOM contains many artefacts, which we aim to remove by re-deriving the solution in-flight.

\subsubsection{MRS ground-based extended fringe flats versus the PSFF}
Assuming the detector electronic effects are well calibrated in the \textit{JWST} \texttt{calwebb\_detector1} module of the pipeline (Morrison et al., in prep.), a subtle yet fundamental aspect of applying a set of PHOTOM files is that the best results are produced when observing a star with (a) the same number of dithers as the absolute flux standard that was used to derive PHOTOM, (b) the same aperture size and aperture correction, and (c) the same fringe correction. Applying a different fringe correction means that a new set of PHOTOM files needs to be derived as the fringes and PHOTOM are linked. 

\begin{figure}
    \centering
    \includegraphics[width=\columnwidth]{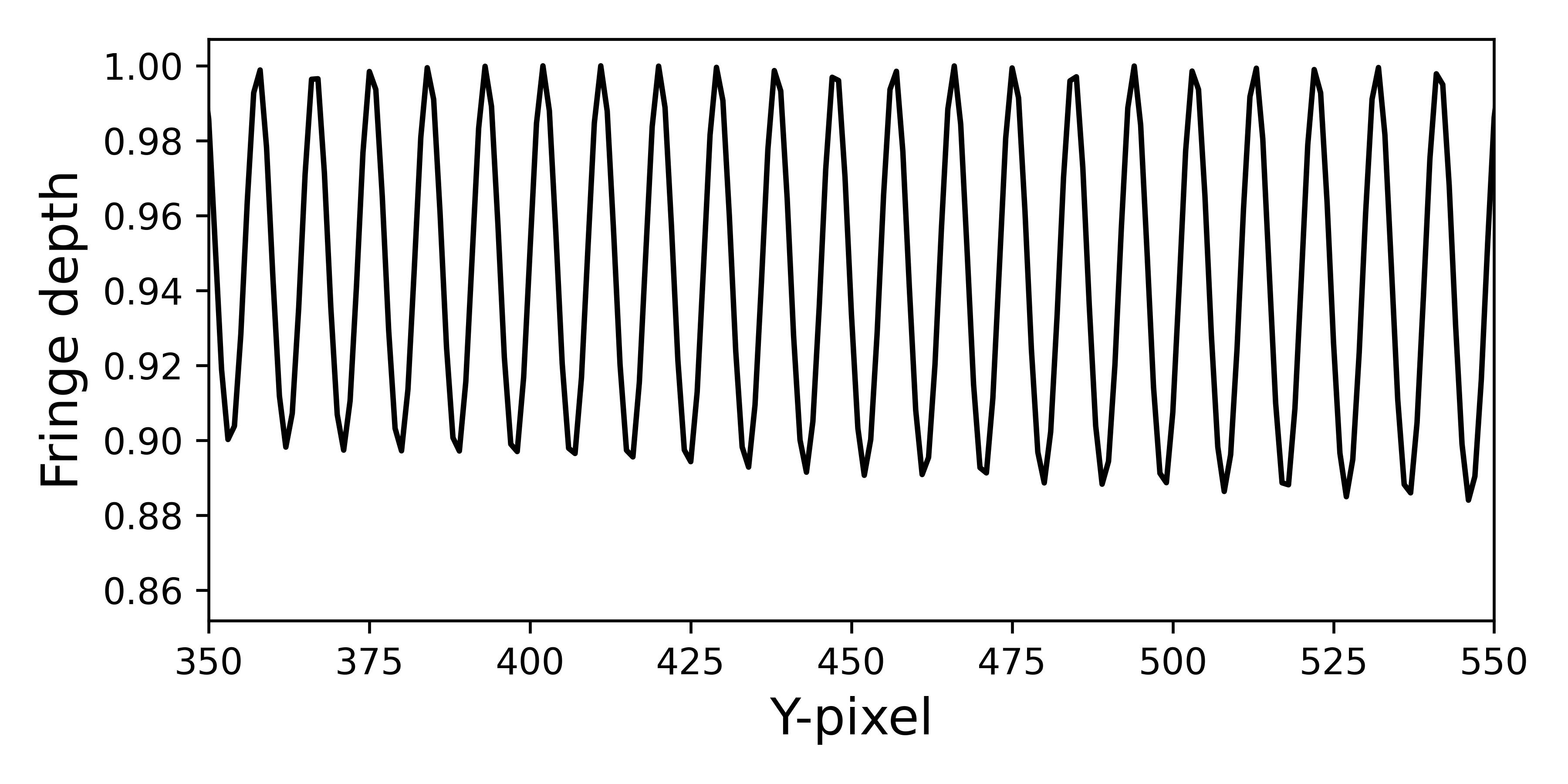}
    \caption{Depth of the extended commissioning fringe flat in band 1A. The values on the y-axis are the correction factors by which the data is divided in the pipeline.}
    \label{fig:fringe_flat}
\end{figure}

The extended fringe flats incorporated in the \textit{JWST} pipeline assume that the peaks of the fringes are the true signal; an example is shown in Fig. \ref{fig:fringe_flat}, where a transmission value of 1 is set at the location of the fringe peaks. The uncalibrated data are divided by this extended fringe flat, bringing the entire spectrum `up' at the fringe peaks, assuming this to be the `true' flux. The PSFF derived in Sect.~\ref{subsec:ps_fringe_flat} assumes the running mean of the fringe interferometric profile to be true, and hence it will not be compatible with the existing \textit{JWST} pipeline reference files. Changing the format to be the same as the commissioning pipeline fringe flats is non-trivial as the fringes shift in phase and become a sum of multiple frequency components in channels 3 and 4. Therefore, the simple image shown in Fig. \ref{fig:fringe_flat} becomes increasingly more complex.

\begin{figure*}
    \centering
    \includegraphics[width=\textwidth]{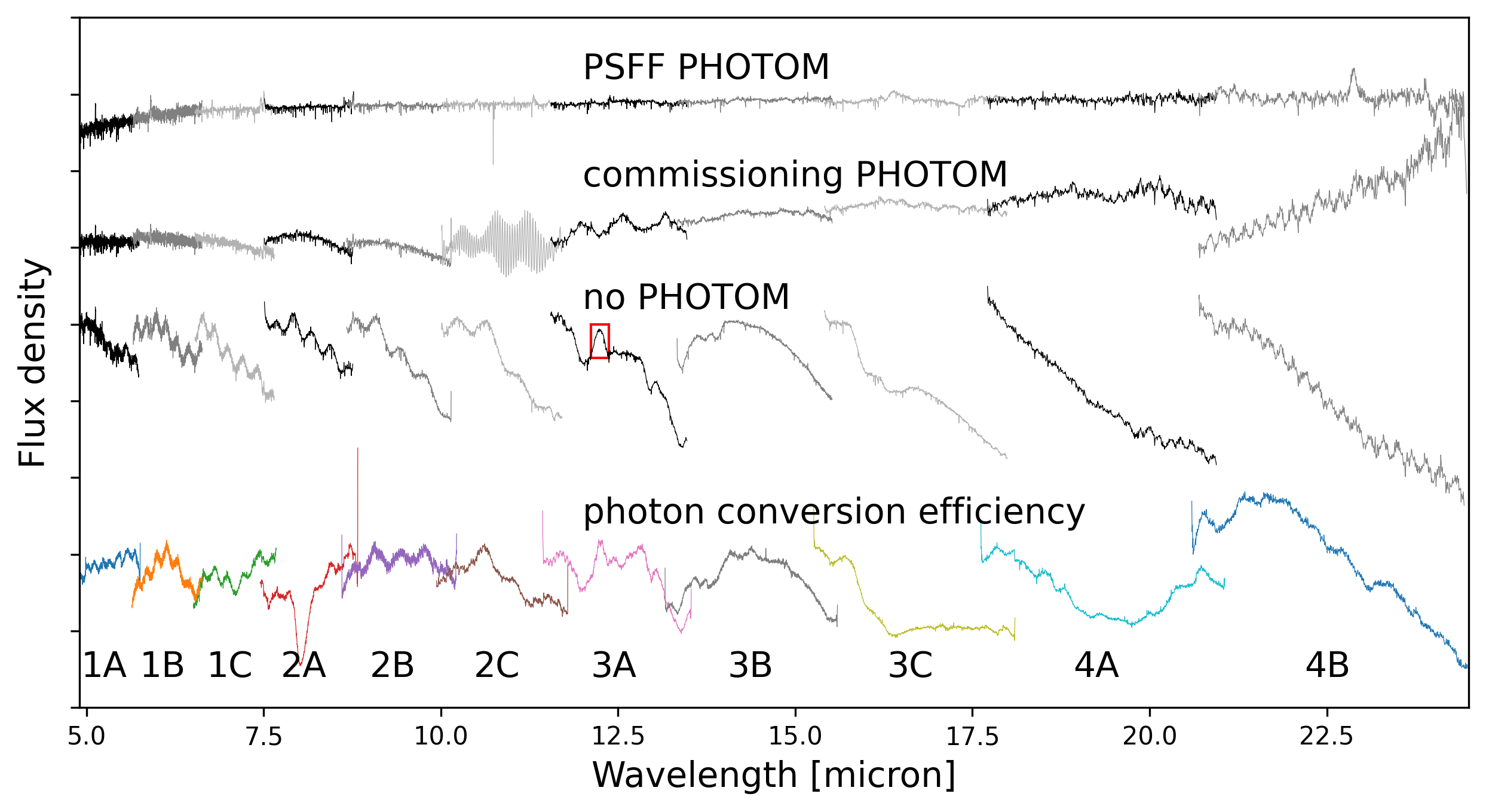}
    \caption{Comparison between the different ways of reducing the data of HD~159222 with the PSFF fringe flat. The curves are (from bottom to top) the photon conversion efficiency of the MIRI MRS detectors; the reduction without spectrophotometric calibration (PHOTOM) applied; the results after applying the commissioning PHOTOM; and the new dedicated PHOTOM. Band 4C is omitted due to low signal. The red box indicates the spectral leak.}
    \label{fig:photom}
\end{figure*}

Figure~\ref{fig:photom} demonstrates the issues discussed. We show four different spectra in 11 of the 12 MRS spectral bands (1A to 4B, excluding 4C due to low S/N). The `photon conversion efficiency' shows the MIRI FM ground testing derived photon conversion efficiency values normalised per band. At 8~$\mu m$ an artefact from the MIRI MTS is visible. The `No PHOTOM' spectra show the extracted spectrum of the G dwarf HD~159222 after reducing the data with a PSFF and without applying the \texttt{photom} step in the \textit{JWST} pipeline. The shape and most of the features correspond to those of the ground spectral response, excluding the 8~$\mu m$ artefact, which is no longer present in the data. The `Commissioning PHOTOM' spectra are those extracted when applying the PSFF, as well as the commissioning version of the PHOTOM reference files. Fringes in band 2C are introduced by the \texttt{photom} step; in this case the fringes are present in the PHOTOM calibration reference file.  Importantly, the change in the choice of fringe flat (pipeline vs new PSFF), with different assumptions about the true signal, results in the PHOTOM files not being applicable, which in turns results in the spectra showing a prominent curvature in each spectral band. The `PSFF PHOTOM'  spectra are extracted after applying the PSFF together with the re-derived PHOTOM reference files based on HD~163466.

\subsubsection{Detector-based background modelling}
Below we describe how we re-derived the PHOTOM reference files based on HD~163466, after self-calibrating by applying the new PSFF fringe flat files derived from HD~163466 itself. Due to the 2D response of the detector, which is incorporated into the spatially extended background signal, the annulus background subtraction will not be accurate here. The aim is therefore to remove the influence of the 2D  response in the background on the detector image plane prior to the spectrophotometric calibration by using the dedicated background observations (see Table \ref{tab:prop}). To minimise the contribution of noise from the dedicated background observations linked to HD~163466, a smooth 2D polynomial is fitted to each dispersed slice of signal on the detector. Since the background has already been subtracted in this case, the spectra are extracted without an annulus. We use different aperture correction factors to account for this. This method was only applied to channels 1-3. For most parts of the wavelength range, the use of the dedicated backgrounds works well. However, the spatial and spectral structure of the stronger thermal background in channel 4 does not match between the dedicated background and the target observation. Due to the large resulting (structured) residuals, a different technique is required to subtract the background.

\subsubsection{Dither-based background subtraction}
\begin{figure}[h!]
    \centering
    \includegraphics[width=0.9\columnwidth]{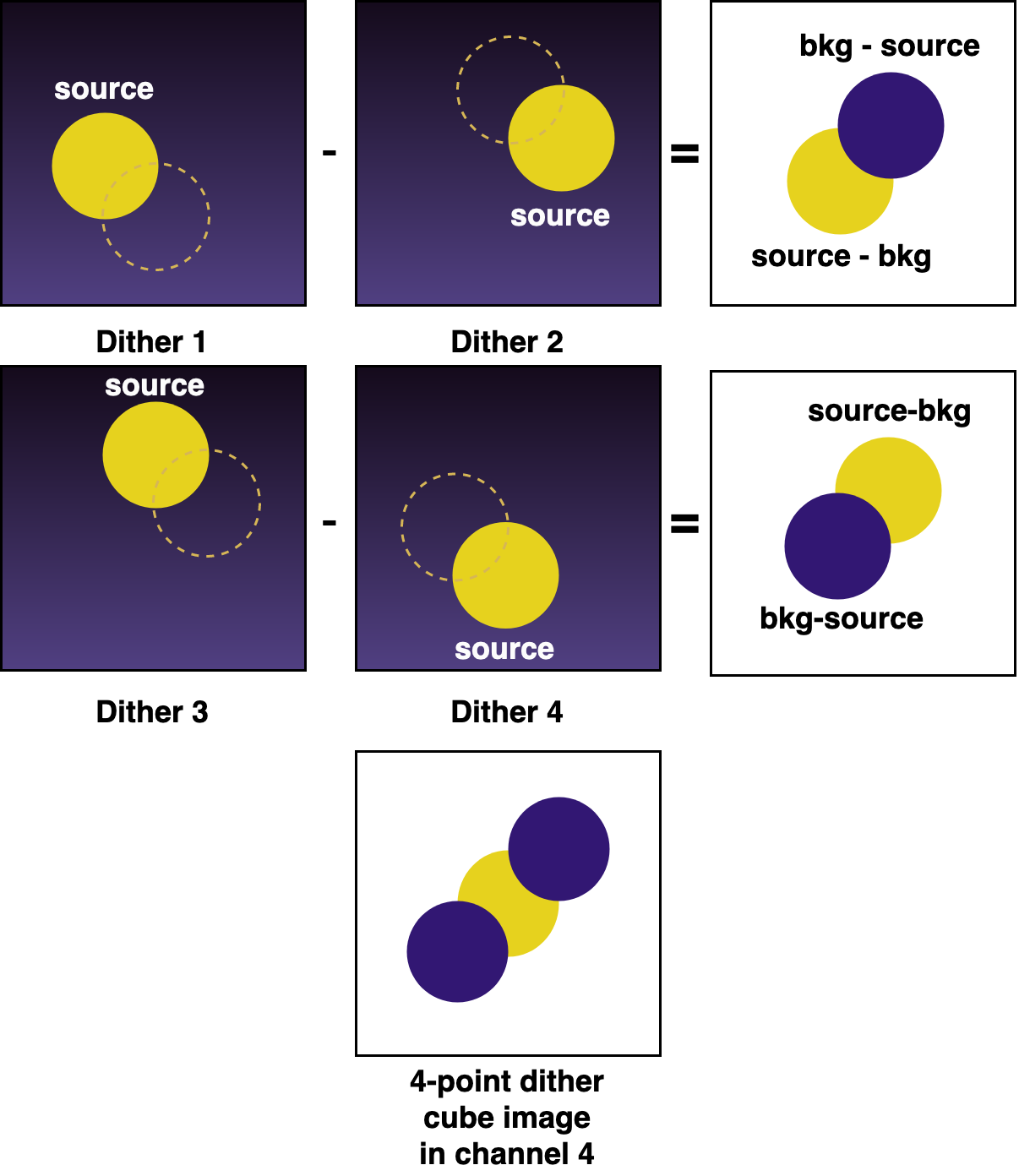}
    \caption{Illustration of the four-point dither subtraction to remove the background. Two pairs of the largest separated dithers (dither 1~--~2, dither 2~--~1; and dither 3~--~4, dither 4~--~3), are subtracted from one another to remove the background. This leaves a clean positive PSF and a negative one. Due to the size of the PSF in channel 4, the negative and positive signal overlap, and therefore a smaller aperture is used to extract the spectrum.}
    \label{fig:dither_subtr}
\end{figure}

To mitigate the issues with using the dedicated background in channel 4, the large drop in quantum efficiency in channel 4 is exploited. By subtracting individual dithers from each other, the background can be decoupled and removed from the signal on the source; this technique is similar to chop-nodding for ground-based infrared telescopes. In MIRI/MRS observations, the purpose of dithering is to better sample the PSF, which is undersampled by design (a four-point dither is recommended). We illustrate the dither-based background subtraction procedure in Fig. \ref{fig:dither_subtr}. Since the signal is not as strong in channel 4 for the A star, the PSF wings are not detected and the resulting negative PSFs do not significantly overlap with the signal. We note that this cannot be done for targets that are bright at the longer wavelengths as the overlap of the PSF wings becomes more significant here. However, to fully mitigate the effect of the negative PSFs, the spectra in channel 4 are extracted with a 1.25$\lambda/D$ aperture instead of 2.5$\lambda/D$. Additionally, it should be noted that due to the low signal in channel 4, the resulting spectra are noisy and the reference files  not as robust as for the other channels. This method is only applied for the PSFF case; all other spectra are extracted using the standard aperture-annulus method described in Sect. \ref{subsec:1d_extract}. 

\subsubsection{Deriving a pointing-dependent ASRF}
\begin{figure}
    \centering
    \includegraphics[width=0.9\columnwidth]{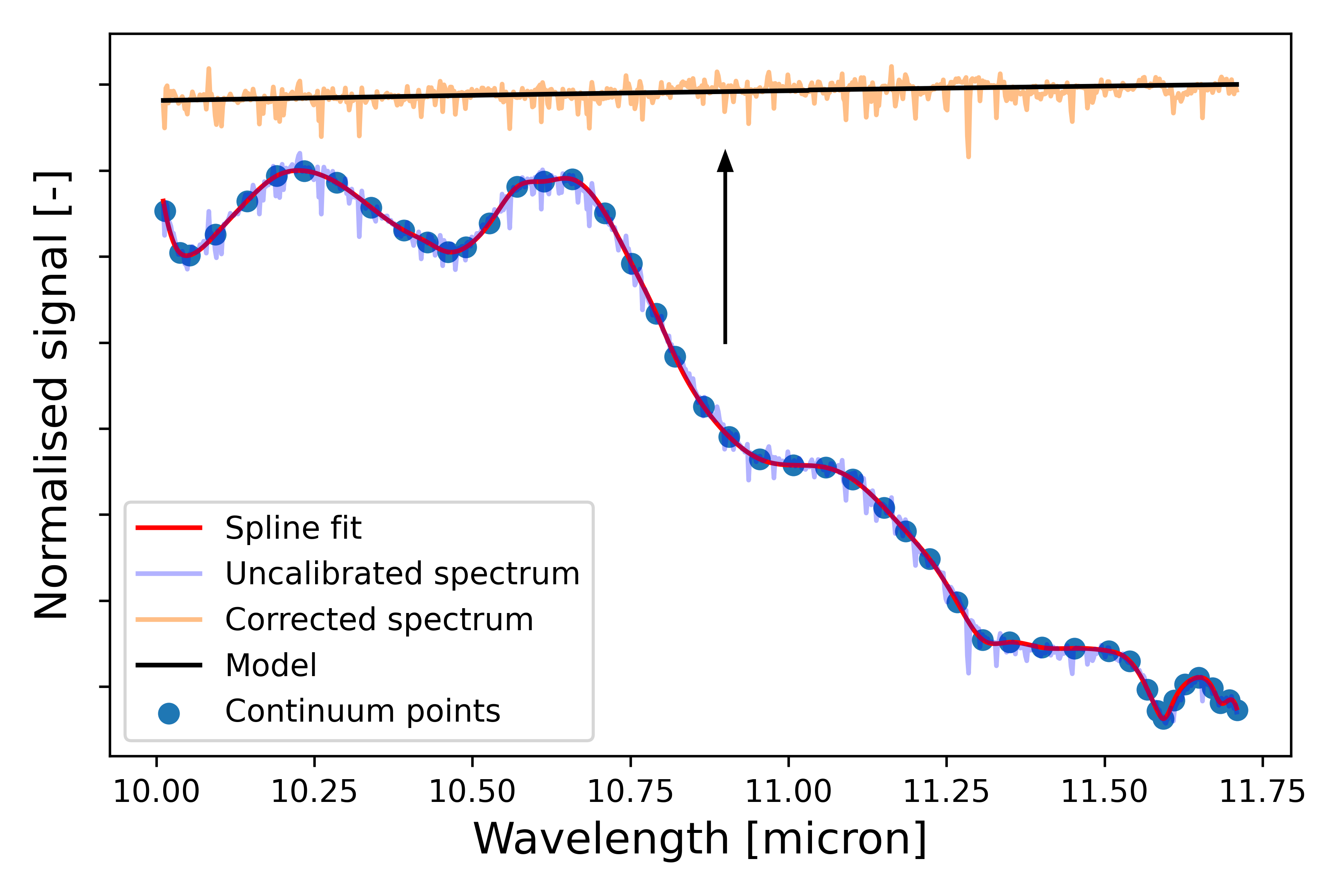}
    \caption{Illustration of derivation process of   the new PHOTOM. This example shows band 2C. The spectrum was normalised with respect to its maximum, and is unit-less. A spline is fitted through the chosen continuum points on the spectrum. This spline is then divided by the model spectrum to find the correction factors. The result from multiplying the uncalibrated spectrum by these factors (indicated by the arrow) is shown in orange, and matches the model.}
    \label{fig:photom_der}
\end{figure}

By defining a spline through hand-selected points on the continuum of HD~163466, the spline-fitted continuum is divided by a model spectrum of the continuum of HD~163466 \citet{ref:17BoMeFl,ref:20BoHuRa}. The precision of PHOTOM is therefore limited by the defined continuum and precision of the assumed model. The resulting wavelength-dependent factors are used as the new PHOTOM correction applied on the G dwarf. This process is illustrated in Fig. \ref{fig:photom_der}. We can safely assume that the systematics in the continuum of the PHOTOM-uncalibrated spectrum are due to the MRS optical chain transmission and the detector response, as presented in Fig. \ref{fig:photom}. The response systematics in the `no PHOTOM' and `photon conversion efficiency' spectra match quite well, aside from the water feature at 8~$\mu m$ (an artefact from the FM ground test campaign).

\subsubsection{Correcting the MRS spectral leak}
\begin{figure}[h!]
    \centering
    \includegraphics[width=\columnwidth]{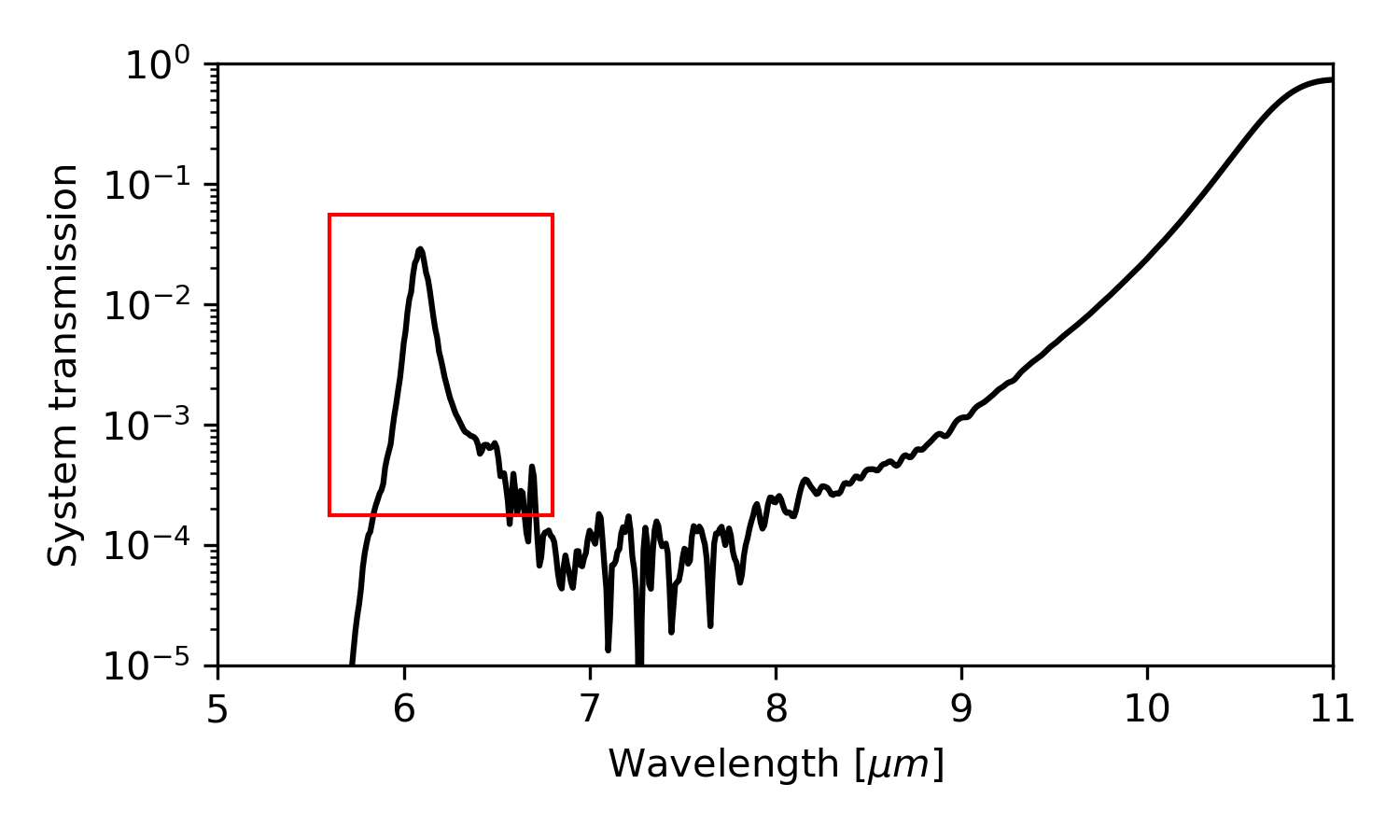}
    \caption{Transmission of the MRS optical path to sub-band 3A. A leak is seen around 6.1~$\mu m$, indicated by the red box.}
    \label{fig:transmission}
\end{figure}

A spectral leak is present in the data of HD~163466, and in fact in all MRS data. In the case of HD~163466 this impacts the PHOTOM solution, and hence it is even more important to address it. The transmission profile of the MRS dichroics causes a spectral leakage of the m=2 grating order being superimposed on the m=1 grating order. Quantitatively, 2.5\% of the spectral flux density at 6.1~$\mu m$ (see system transmission based on lab data of dichroic transmission curves in Fig. \ref{fig:transmission}) is added to the 12.2~$\mu m$ wavelengths (band 3A) (for further details, see \citealt{phdthesisYannis}, Sect. 7.3). Due to the presence of the leak in the reference star, also visible in Fig. \ref{fig:photom}, the increase in flux is folded into the commissioning PHOTOM reference file in band 3A, resulting in an underestimated correction factor around this wavelength. To remove this feature from the newly derived PHOTOM files, we find the fraction of flux from band 1B that is present in 3A using the transmission of the optical path shown in Fig. \ref{fig:transmission}.

Since we know that the flux per band is equal to the signal $S$ times the correction factors $P$, the fluxes $F$ are
\begin{align}
\begin{aligned}
    F_{1B} (\lambda_{1B}) = S_{1B}(\lambda_{1B}) \cdot P_{1B}(\lambda_{1B})\text{,} \\
    F_{3A} (\lambda_{3A}) = S_{3A}(\lambda_{3A}) \cdot P_{3A}(\lambda_{3A})\text{,}
\end{aligned}
\end{align}
and inversely
\begin{align}
\begin{aligned}
    P_{1B} (\lambda_{1B}) = \frac{F_{1B}(\lambda_{1B})}{S_{1B}(\lambda_{1B})}\text{,} \\
    P_{3A} (\lambda_{3A}) = \frac{F_{3A}(\lambda_{3A})}{S_{3A}(\lambda_{3A})}\text{.}
\end{aligned}
\end{align}
The flux in the leak $L$ is related to the transmission $T_{sys}$ as
\begin{equation}
    L_{1B} (\lambda_{1B}) = F_{1B}(\lambda_{1B}) \cdot T_{sys}(\lambda_{1B}) = S_{1B}(\lambda_{1B})\cdot  P_{1B}(\lambda_{1B}) \cdot T_{sys}(\lambda_{1B}) \text{.}
\end{equation}
Since the spectral leak in the A star results in a positive bump in the spectrum, the PHOTOM factors are underestimated by a factor
\begin{align}
\begin{aligned}
    P_{3A,corr} = \frac{[L_{1B} (\lambda_{1B})](\lambda_{3A})}{S_{3A}(\lambda_{3A})} \\ = \frac{[S_{1B}(\lambda_{1B})\cdot  P_{1B}(\lambda_{1B}) \cdot T_{sys}(\lambda_{1B})](\lambda_{3A})}{S_{3A}(\lambda_{3A})} \text{,}
\end{aligned}
\end{align}
which are added to the PHOTOM factors found from fitting the spline to the continuum:
\begin{equation}
    P_{3A,new} = P_{3A,old} + P_{3A,corr} \text{.}
\end{equation}
In addition to removing the spectral leak artefact from the reference files, the leak from the target itself must also be removed from the spectrum. Following the relations mentioned above, the extra flux is removed using
\begin{equation}
    F_{3A,new} (\lambda_{3A}) = F_{3A,old} (\lambda_{3A})-[L_{1B} (\lambda_{1B})](\lambda_{3A}) \text{.}
\end{equation}

The new spectrum is shown at the top of Fig. \ref{fig:photom}. We note that the system transmission profile shown in Fig. \ref{fig:transmission} had to be shifted by $\sim$0.02 $\mu$m. The lab conditions were likely not representative of in-flight conditions, causing this change.


\section{Results}
\label{sec:results}

\subsection{Propagated noise from the PSFF fringe flat}
\label{subsec:noise}
Since no filtering was performed to create the PSFF, noise is propagated to the G dwarf spectrum when applying the correction. In order to examine the extent of this effect, we find the standard deviation of the noise around the continuum. The smaller the standard deviation, the less noisy the spectrum is. Due to a known issue with the MRS pointing repeatability (Argyriou et al. in prep., and references therein), there is an offset between the observations of the G and A stars in channel 1 on the detector. Because of this offset the PSF is sampled differently. This results in a change in the resulting fringe pattern (see Sect. \ref{subsec:ps_fringe_flat}), and therefore the correction is not expected to work as well. While a pointing offset would compromise the fringe correction in all channels, the requirement is especially stringent in channel 1 due to undersampling of the PSF. The PHOENIX model \citep{ref:17BoMeFl,ref:20BoHuRa} shows a dense forest of features in this wavelength range, making it difficult to distinguish noise from real features. We   therefore examine the noise in channel 2 and onward.

\begin{table}[h!]
\centering
\caption{Standard deviation of the noise after removing the continuum. The values correspond to the two pipeline approaches, the extended fringe flat including the residual fringe correction (left) and the PSFF fringe flat (right).}
\label{tab:noise}
\begin{tabular}{ll}
\textbf{Band} & $\mathbf{\sigma}$ (ff+rfc / PSFF) \\ \hline
2A            & 0.0082 / 0.0086                         \\
2B            & 0.0068 / 0.0071                         \\
2C            & 0.0076 / 0.0076                         \\
3A            & 0.0075 / 0.0064                        \\
3B            & 0.0066 / 0.0057                        \\
3C            & 0.0070 / 0.0069                        
\end{tabular}
\end{table}

The properties of the noise are presented in Table \ref{tab:noise}. In terms of noise the performance of the PSFF is comparable to that of the pipeline fringe flat (based on a spatially homogeneous and extended source) plus the a residual fringe correction.

\subsection{Performance of fringe removal}
\label{subsec:fringe_perf}
\begin{figure}[h!]
    \centering
    \includegraphics[width=\columnwidth]{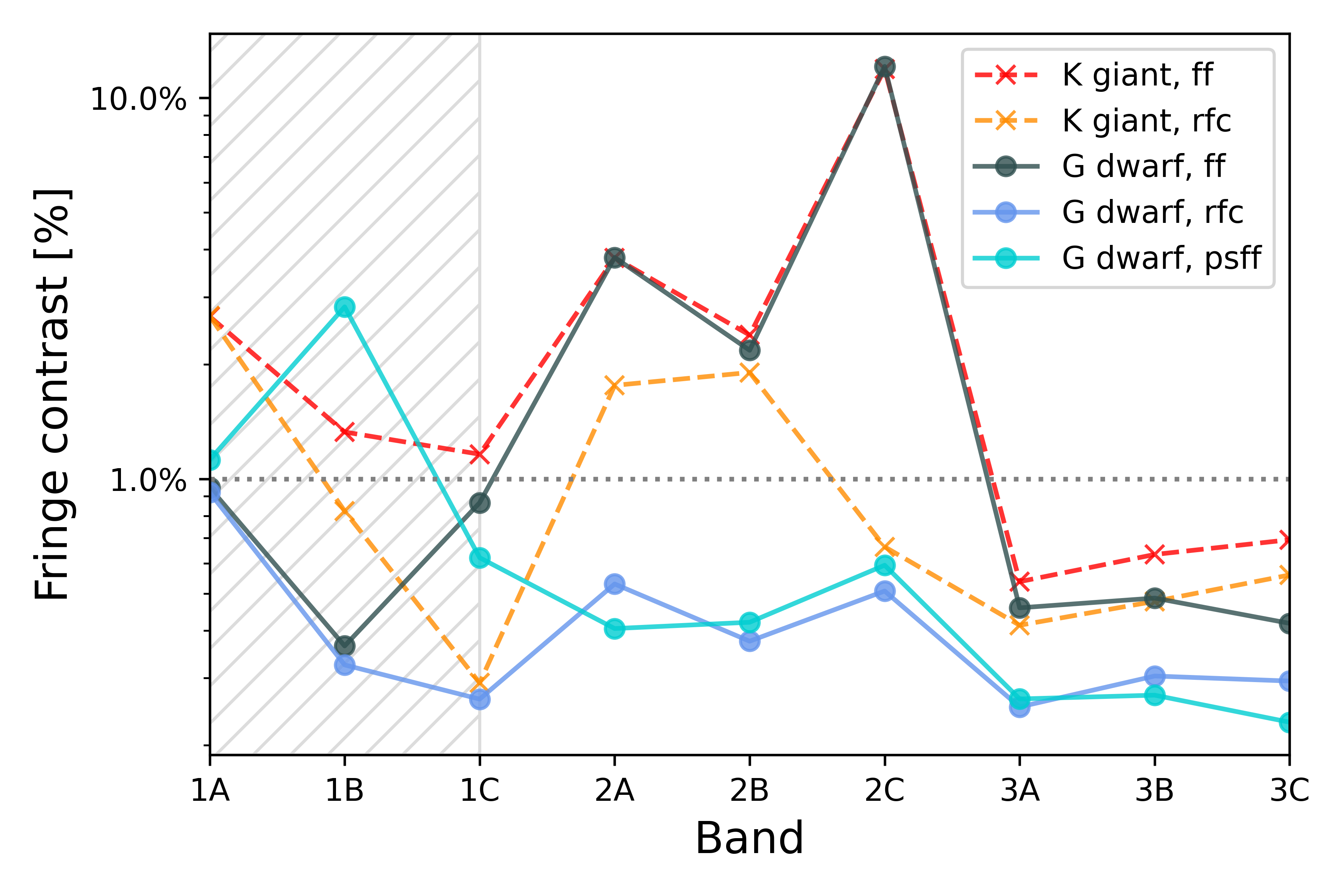}
    \caption{Fringe contrast after the different corrections presented in this work. The curves are colour-coded (see inset): `ff' stands for extended fringe flat, `rfc' for extended fringe flat+residual fringe correction applied,  and `psff' denotes the PSFF. The contrasts in channel 1 are shaded since the pointing repeatability issue results in larger residuals in the PSFF case.}
    \label{fig:contrast}
\end{figure}

When the fringes are removed by the extended fringe flat and residual fringe correction, no components linked to the detector geometric and refractive properties should remain, and no new frequencies should be added. From the estimated residual fringe contrast we can find how the different reduction methods compare. This contrast is approximated based on twice the amplitude of a sinusoid fit to the band, divided by the spectral baseline. The results of this process are presented in Fig.~\ref{fig:contrast}.

We already noted the pointing issues in channel 1, and this is confirmed by the larger fringe residual after applying the PSFF. When looking at the longer wavelength bands, where we  expect a relatively good correction, it is evident that the performance is similar to the extended fringe flat plus residual fringe correction, reaching sub-percent level residuals for the G dwarf. Therefore, the PSFF method results in comparable fringe residuals.

On the other hand, the K giant clearly shows larger contrasts, even after applying the residual fringe correction, particularly in bands 1A and 2A/B. This shows that these results are not only pointing-specific, but also target-specific. The K giant is dense in features, as we  demonstrate in Sect. \ref{subsec:features}, and the residual fringe correction may either not affect certain sections, or potentially remove or reduce the strength of molecular bands. This can be a risk if the periodicities of the molecular bands and the fringes overlap. Additionally, this   results in a larger amplitude of the fitted sinusoid, and therefore a larger estimated fringe residual, as visible in the K giant. Looking at the amplitude of the molecular bands in Fig. \ref{fig:spectra_zoom}, depending on the molecule, this could easily be 10\%, or even larger. We take a closer look at this issue by discussing the periodograms of the stellar and synthetic spectra.

\begin{figure}[h!]
    \centering
    \includegraphics[width=\columnwidth]{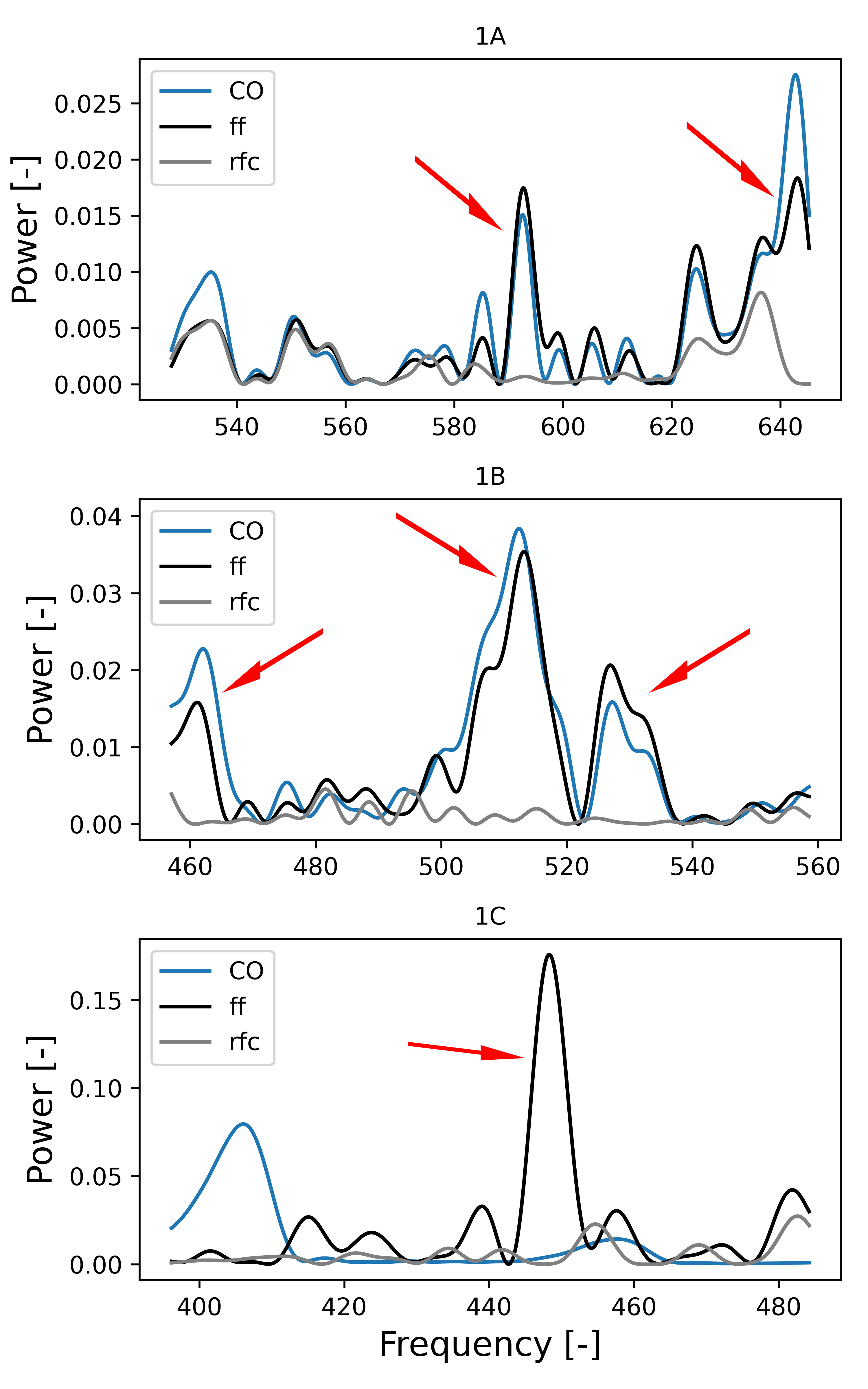}
    \caption{Periodograms of the synthetic CO spectrum and the K giant spectrum per MRS band, before and after residual fringe correction. The curves are colour-coded (see insets): `ff' stands for extended fringe flat and `rfc' for extended fringe flat+residual fringe correction. The frequency range shown is the range where the \texttt{residual\_fringe} pipeline algorithm searches for frequencies to remove. The red arrows indicate the largest peaks that are removed by the algorithm.}
    \label{fig:periods}
\end{figure}

In Fig. \ref{fig:periods} we present the periodograms generated using \texttt{LombScargle} of \texttt{astropy} \citep{astropy:2013,astropy:2018}. The periodograms of the K giant spectrum after applying only the pipeline fringe flat and after applying the additional residual fringe correction, are shown alongside the periodogram of the CO synthetic spectrum. Comparing the profiles, it can be concluded that the CO signal in the spectrum has been directly affected by the residual fringe correction in the cases of bands 1A and 1B, but not   1C. The largest peaks removed by the algorithm are indicated by the red arrows. These overlap with strong peaks in the periodogram of CO in bands 1A and 1B. These peaks are present in the spectrum with only the extended fringe flat applied, and removed after applying the residual fringe correction. In 1C, a detector fringe seems to be removed instead. Therefore, in addition to performing similarly in terms of fringe residuals, using the PSFF mitigates the risk of removing molecular features.

This exercise can be done for all examined MRS bands where the molecular species show prominent features, and rather than showing the periodograms separately, we take the difference between before and after applying the residual fringe correction to demonstrate where the largest differences occur. If this change overlaps with a peak in the molecular periodogram, it is likely that real features are affected rather than solely the detector fringes. The results are included in Fig. \ref{fig:diff_periodogram} in Appendix \ref{app:apppendixA}.

\subsection{Detectability of molecular features}
\label{subsec:features}

\begin{figure*}
    \centering
    \includegraphics[width=0.95\textwidth]{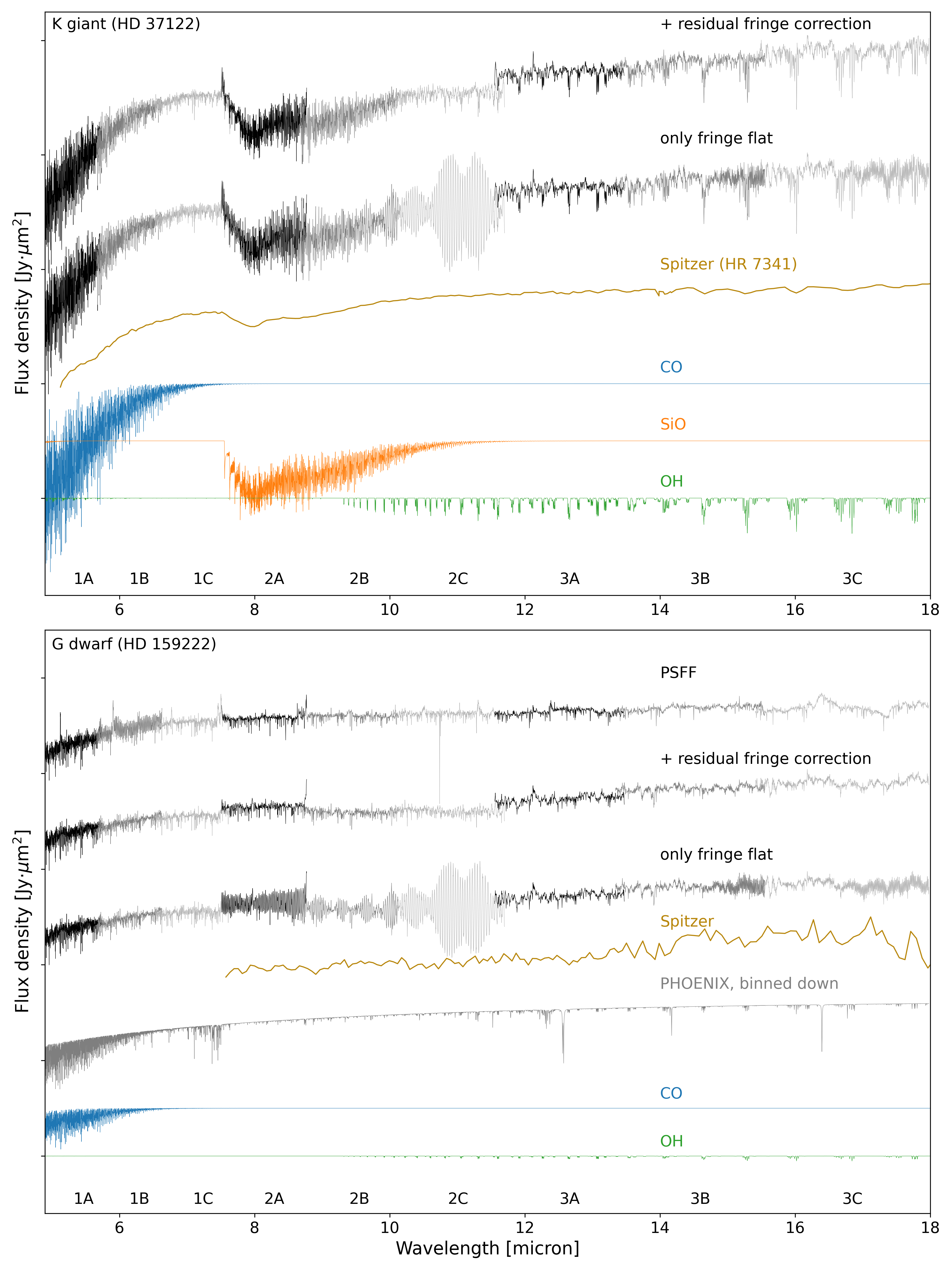}
    \caption{Spectra for HD~37122 (top) and HD~159222 (bottom), after applying only the extended fringe flat, extended fringe flat and residual fringe correction, or the point source specific correction labelled "PSFF"). The y-axis is in Rayleigh-Jeans units, resulting in a flat tail of the stellar spectrum. The sub-bands (A, B, C) are indicated by increasingly light colours; blue, orange, and green are the model CO, SiO, and OH spectra, respectively. The \textit{Spitzer} spectra are taken from \citet{ref:15SlHeCh} (top panel) and the \textit{CASSIS} database \citep{ref:11LeBaSp} (bottom panel).}
    \label{fig:spectra}
\end{figure*}

\begin{figure*}
    \centering
    \includegraphics[width=\textwidth]{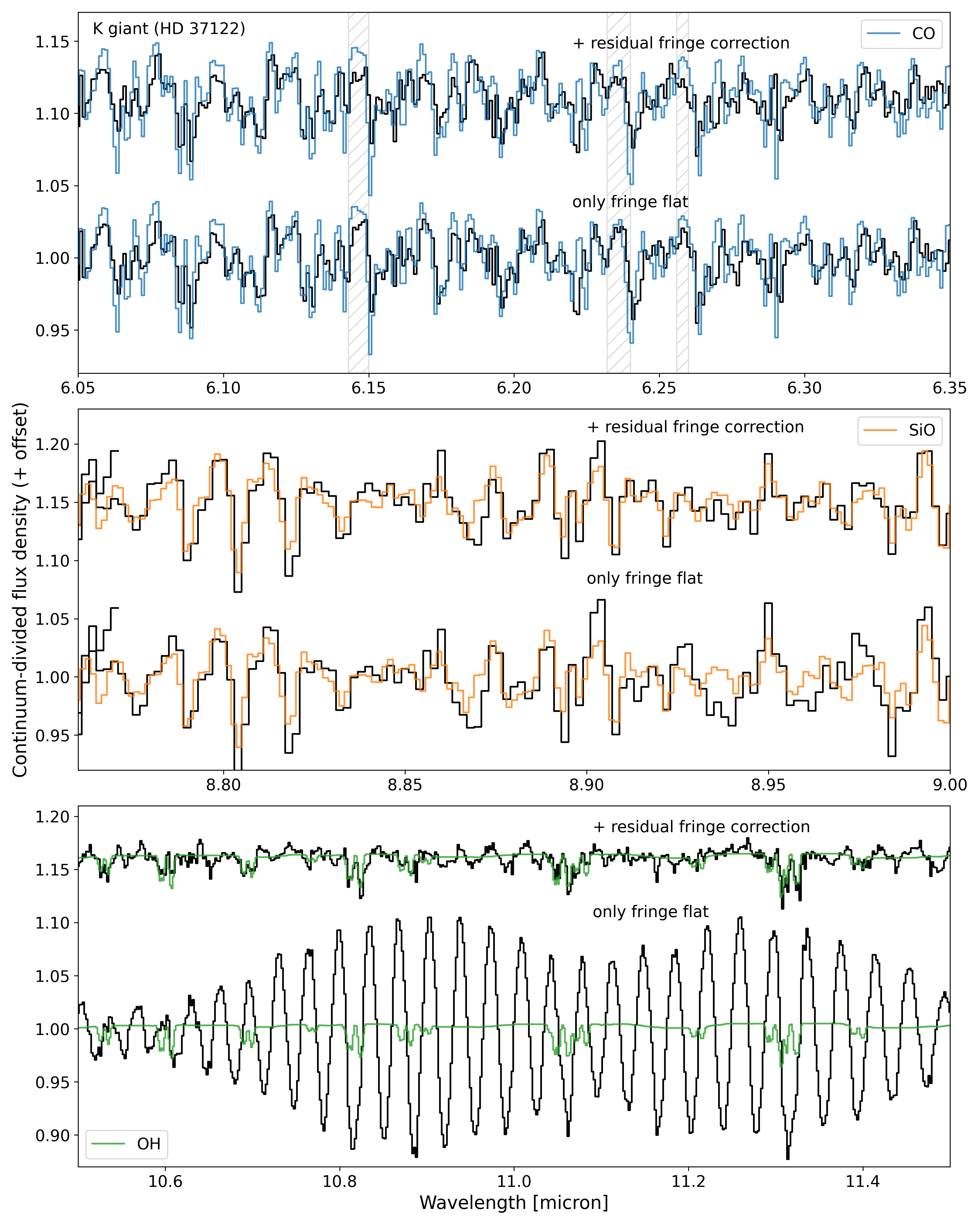}
    \caption{Close-ups of the spectra of HD~37122, with the continuum removed. The bottom spectrum in each plot includes only the extended fringe flat, and the top spectrum both the extended fringe flat and the residual fringe correction. CO, SiO, and OH are overplotted. Bands 1B, 2B, and 2C are shown (from top to bottom). In the top panel some areas where the residual fringe correction reduces the match with the synthetic CO are highlighted.}
    \label{fig:spectra_zoom}
\end{figure*}

\begin{figure*}
    \centering
    \includegraphics[width=\textwidth]{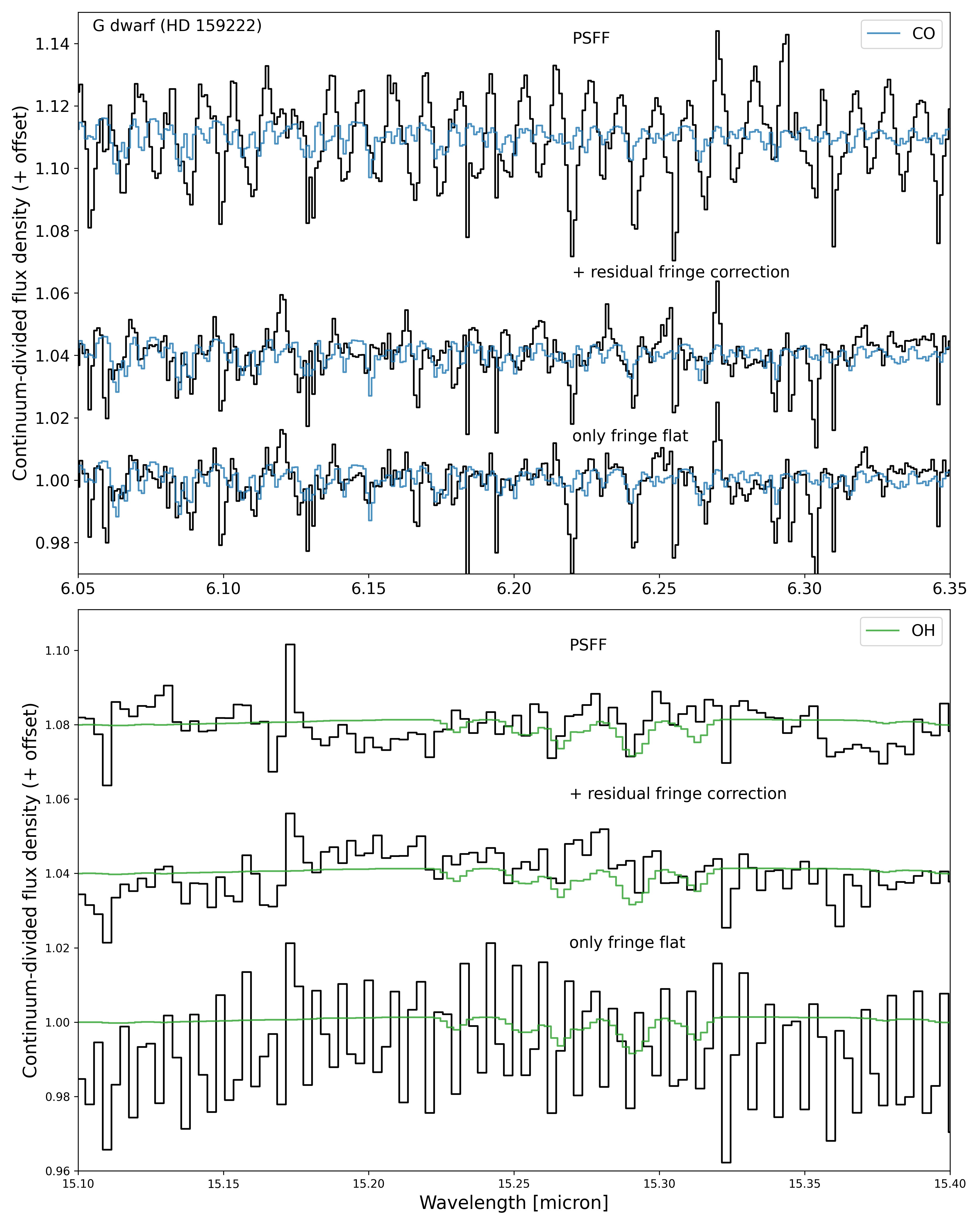}
    \caption{Close-ups of the spectra of HD~159222, with   continuum removed. The bottom spectrum includes only the extended fringe flat, the middle both the extended fringe flat and residual fringe correction, and the top  the point source fringe flat. CO and OH are overplotted. Bands 1B (top panel) and  3B (bottom panel) are shown.}
    \label{fig:spectra_zoom_hd159}
\end{figure*}

The resulting normalised spectra for both the G dwarf and the K giant are presented in Fig.~\ref{fig:spectra}. The top panel of Fig.~\ref{fig:spectra} shows the K giant, the bottom panel the G dwarf. Figures  \ref{fig:spectra_zoom} and  \ref{fig:spectra_zoom_hd159} include closer comparisons for the K giant of the continuum-divided stellar spectra and the synthetic spectra of CO, SiO, and OH.

While previously more detailed statistics regarding the corrections were discussed, these spectra in Fig.~\ref{fig:spectra} demonstrate the global effects of the different methods on the spectra directly. In band 2C the commissioning PHOTOM files introduced fringes, which are subsequently removed by the residual fringe correction. The aim is to fix this issue during Cycle 1. Additionally, the spectra agree much better for the PSFF method compared to the pipeline output, where small jumps are present between bands.

In the last section the risk of the residual fringe correction removing or altering features was mentioned. A direct example of this effect is visible in the top panel of Fig. \ref{fig:spectra_zoom}, where the highlighted areas indicate the features are altered and match the CO synthetic spectrum less well. However, the positive impact is also visible. For the K giant this is best demonstrated in the third panel of Fig. \ref{fig:spectra_zoom}, where the large fringes initially masked OH features. At the shorter wavelengths, particularly in the spectrum of the K giant, the signal seems noisy. This is due to the presence of CO, as shown in Fig. \ref{fig:spectra_zoom}. In addition to the CO bands, the SiO dip around 8~$\mu m$ is clearly visible in Fig.~\ref{fig:spectra}. While the \textit{Spitzer} IRS spectrum already shows the drop in flux around 5--6~$\mu m$ due to CO, the dip from SiO around 8--10~$\mu m$, and the OH features longward of 14~$\mu m$, the detailed views of the MRS spectrum in Fig. \ref{fig:spectra_zoom} demonstrate how well we are now able to observe the structure in the spectrum of the K giant, which  matches the synthetic spectra very well.

On the other hand, the G dwarf is less rich in features, due to its higher effective temperature, evident in the flat CO spectrum in Fig. \ref{fig:spectra_zoom_hd159}. The features of SiO and OH are not confidently detected in this temperature range (see Fig. \ref{fig:spectra}). While similar improvements are observed between the extended fringe flat and addition of the residual fringe correction, in the topmost spectrum of the bottom panel of Fig. \ref{fig:spectra}, only the PSFF was applied to the data. The previous sections discussed that the statistics demonstrate that the correction performs well, but here the visual effect on the spectrum is shown. It is evident that, with just one fringe correction, a clean spectrum is achieved. However, the caveat related to the pointing offset in channel 1 is also visible as residual fringing, especially in the top panel of Fig. \ref{fig:spectra_zoom_hd159}. To show an area where vast improvements are seen, the bottom panel is included, where the single correction   greatly reduces the fringe contrast.

\begin{figure}[h!]
    \centering
    \includegraphics[width=\columnwidth]{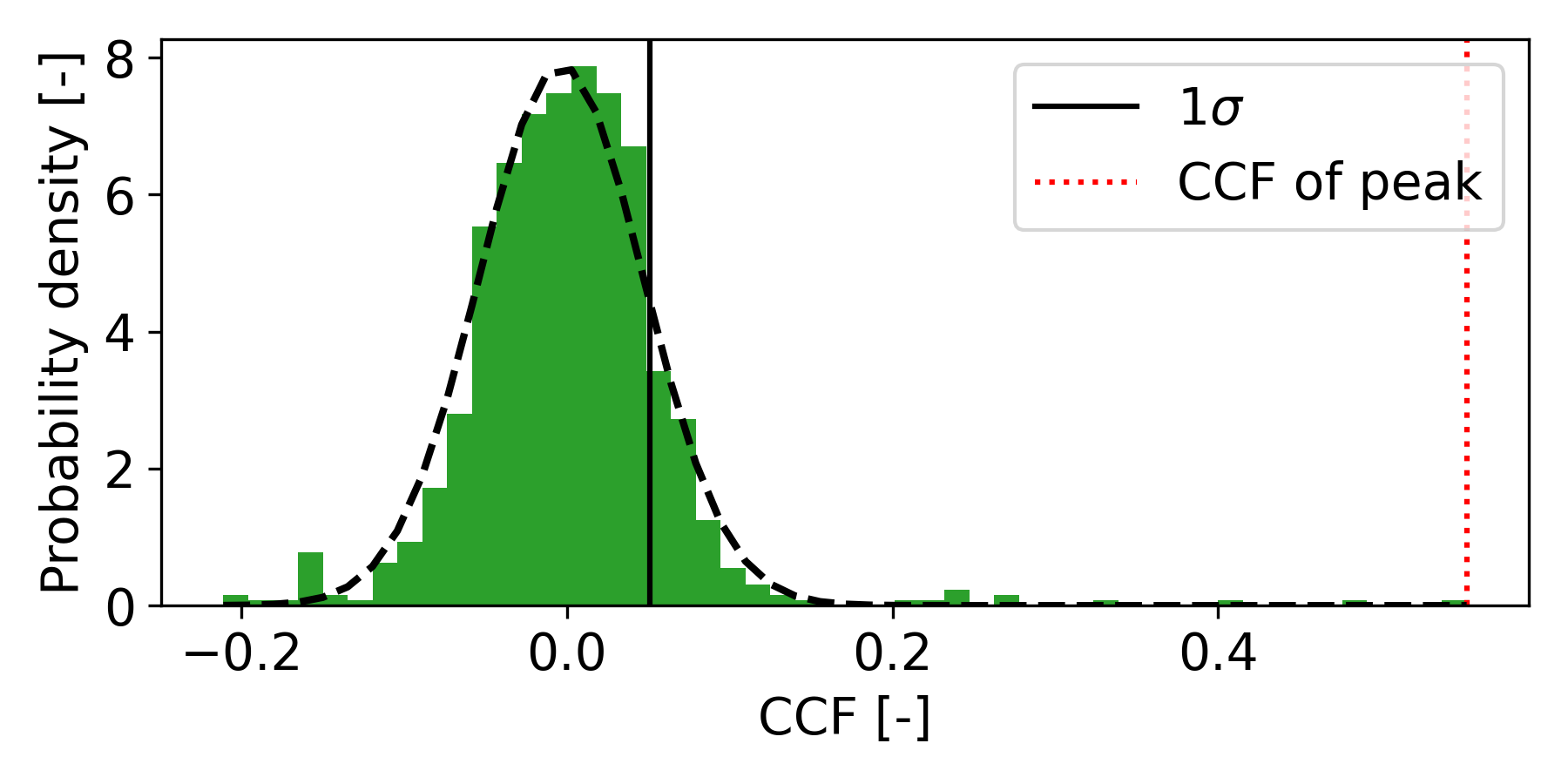}
    \caption{Example of cross-correlation histogram of OH in band 2C. To find the S/N, the peak cross correlation value (CCF) is divided by the estimated standard deviation of the noise ($1\sigma$).}
    \label{fig:cross_corr}
\end{figure}

In order to better quantify improvements in the data reduction and strength of the features in the spectra, we perform cross-correlations with the synthetic spectra discussed in Sect. \ref{subsec:data} and find the resulting S/N of the detection. First, the spectra are binned to the same resolution as the observations, using \texttt{SpectRes} \citep{ref:17Ca}. The continuum of both the synthetic spectra and the observations is removed using a running mean, such that only the structure of the molecular bands is correlated. The cross-correlation is done with the standard \texttt{numpy.correlate} function. We find the S/N similarly to \citet{ref:21PeBoCh} and \citet{ref:22PaNaCu}: the maximum correlation factor is divided by the noise estimated from the standard deviation of a Gaussian fitted to a histogram of the correlation results. An example of this is illustrated in Fig. \ref{fig:cross_corr}. We use the data from an entire sub-band, but only the bands where the molecular features cover the entire band. However, we note that since the fringes in the spectrum are not random in nature, the Gaussian approximation of the noise may not be entirely accurate in cases where large fringe residuals remain.

The S/N values are summarised in Table \ref{tab:cross_corr}, where the results are only shown when a molecule covers the entire band, and significant S/N is found in at least one of the different data reduction versions. The location of the cross-correlation factor peaks is consistent between the different bands, and centred. In some bands the pipeline fringe flats leave large residuals, and the cross-correlations are not able to find the molecular signal in the spectrum. For example, in the extreme case of band 2C of the K giant HD~37122 the S/N of the OH detection drastically increases after running the residual fringe step. However, the results of both reductions are similar, showing that generally the presence of some residual fringes does not prevent the detection of these relatively strong features. In   the K giant all three of the molecules are detected, with S/N values near 10 or higher.

Interestingly, in some cases the S/N is reduced after applying the residual fringe correction. This is most significant in the CO detections in bands 1A and 1B, and OH in 3A. In Sect. \ref{subsec:fringe_perf} it was demonstrated that some part of the periodogram of CO overlapped with the expected fringe frequencies. Peaks in the same location as those of CO were removed by the residual fringe correction in 1A and 1B while remaining untouched in 1C, indicating that the nature of the CO features was altered by the algorithm in these cases. It is therefore possible that the reduced S/N is due to this change, and indeed the observed spectrum seems to match less well   the synthetic spectrum in Fig. \ref{fig:spectra_zoom}, where some affected areas are highlighted by a hatched pattern (for example before 6.15~$\mu m$, where a bump is removed from the spectrum between the extended fringe flat and residual fringe correction). These results also show why the S/N is increased in other cases; in bands 1C and 2A--C in Fig. \ref{fig:diff_periodogram} the largest difference between the two corrections occurs in locations where no strong molecular peaks are found. While the effect in the demonstrated cases is relatively small (detection was not prevented by the residual fringe correction) the impact and limitations of the algorithm must be considered when examining spectra.

In the G dwarf a significant S/N is found for CO only in bands 1A and 1B. Due to the higher stellar effective temperature, the CO features are weaker, and quickly disappear into the noise. Due to the performance of the PSFF in channel 1, a decrease in S/N is observed compared to the other two reduction methods. The same issue is seen here as in the K giant, where the S/N decreases between extended fringe flat and residual fringe correction. This is also visible in the match between the synthetic spectrum and the different spectra in Fig. \ref{fig:spectra_zoom_hd159}.

\begin{table*}[h!]
\centering
\caption{Signal-to-noise ratios of the cross-correlations between the synthetic molecular spectra and the different spectral bands. The first number corresponds to the data without the residual fringe correction, while the second includes the extra residual fringe correction step. The third value in the G dwarf rows denotes the cross-correlation from the point source fringe flat. Entries flagged with a dash indicate that there are no significant features in this band or that no significant S/N was found.}
\label{tab:cross_corr}
\begin{tabular}{llllllllll}
    & \multicolumn{9}{c}{\textbf{K~giant (HD~37122)}}                                                                                                                                                                                         \\ \hline \hline
    & \multicolumn{1}{c}{1A} & \multicolumn{1}{c}{1B} & \multicolumn{1}{c}{1C} & \multicolumn{1}{c}{2A} & \multicolumn{1}{c}{2B} & \multicolumn{1}{c}{2C} & \multicolumn{1}{c}{3A} & \multicolumn{1}{c}{3B} & \multicolumn{1}{c}{3C} \\ \hline
CO  & 19.6 / 19.1            & 15.7 / 13.6            & 6.9 / 7.8             & -              & -              & -              & -              & -              & -              \\
SiO & -              & -              & -              & 8.8 / 9.4              & 10.9 / 11.3           & 3.7 / 5.2              & -              & -              & -              \\
OH  & -              & -              & -              & -              & -              & 3.7 / 10.9              & 8.5 / 7.9              & 10.9 / 10.1              & 11.5 / 11.5              \\ \hline \hline
    & \multicolumn{9}{c}{\textbf{G~dwarf (HD~159222)}}                                                                                                                                                                                         \\ \hline \hline
CO  & 14.2 / 13.7 / 13.4            & 9.0 / 7.5 / 4.0            & -              & -              & -              & -              & -              & -              & -              \\
OH  & -              & -              & -              & -              & -              & -              & -              & -             & -          
\end{tabular}
\end{table*}


\section{Discussion}
\label{sec:discussion}

\begin{figure*}
    \centering
    \includegraphics[width=0.9\textwidth]{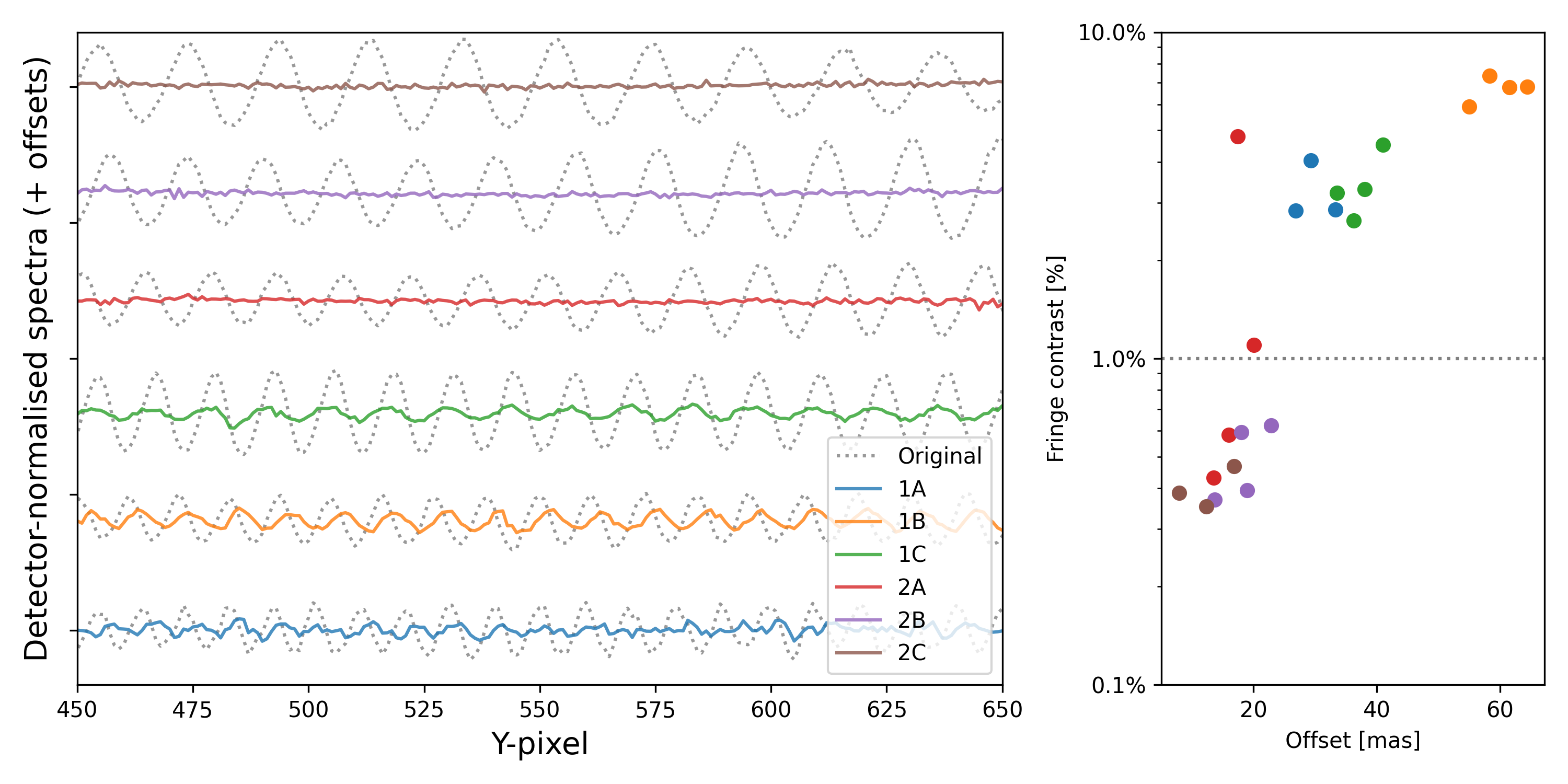}
    \caption{Fringe residuals on the detector of the G dwarf data after applying the PSFF correction for the different sub-bands of channel 1 and 2 shown as the profile (left) and the fringe contrast (right). The offset in the right panel is measured compared to the location of the A star. The larger this offset, the larger the fringe residual.}
    \label{fig:contr_off}
\end{figure*}

We found the PSFF to work most effectively in channels 2 and 3, and less well in channel 1 (due to the MRS pointing non-repeatability), with relatively high remaining fringe contrast especially in bands 1B and 1C (see Fig. \ref{fig:contrast}). Channel 1 has the most stringent requirement on accuracy of target acquisition for the PSFF due to the MRS PSF being spatially undersampled by design. It remains to be seen to what extent we will be able to improve our ability to use empirical PSFFs in channel 1, where the pointing offset resulted in percent-level residuals, rather than sub-percent (see Fig. \ref{fig:contrast}). Calibration observations sampling a sub-pixel raster scan would be the most straightforward approach to get representative fringe flat. Currently, there is not enough data to constrain how large the pointing offset may be for the method to work correctly, though the data from the different dithers of the G dwarf in Fig. \ref{fig:contr_off} suggest that the largest increase in error occurs for an offset of 20--33 mas. Nevertheless,   using target acquisition is necessary to apply this method. Especially when considering features potentially being removed by the residual fringe correction, or altered as  shown here, the PSFF should still be the preferred method for fringe removal in the future in channels 2, 3, and 4, and potentially in channel 1 once a larger sample of pointing-specific PSFFs is available.

In addition to the risk of the period of the real spectral features overlapping with those of the detector fringes, filtering out a sinusoid can still indirectly change other parts of the spectrum. This, and the above-mentioned problem where the residual fringe algorithm directly changes parts of a molecular feature, is likely not stringent for most applications, but can become important when searching for fainter features. Additionally, feature-rich spectra may not be corrected well by the residual fringe correction, in which case the PSFF is the preferred method.

Care must be taken when deriving a PSFF. We  touch upon the difficulties and consequences of this process in Sect.~\ref{subsec:photom}. For a perfect correction any spectral lines must be removed, and the reference data must be filtered to reduce noise, which would otherwise be propagated into the spectrum. Although in Sect. \ref{subsec:noise} we did not find any significant increases in the noise levels, it is still advised to find a way to remove the noise. We did not do this here, but we aim to improve our methods for future work.

Finally, the commissioning PHOTOM files have some issues, for example re-introducing fringes in band 2C (see Sect. \ref{subsec:photom}). In future versions these issues will be worked out by deriving the reference files from scratch in-flight. By doing so, features introduced in the PHOTOM files long ago will no longer be present. We already presented a method here to do this specifically for point sources in a particular dither pattern, but we must fully probe the 2D response of the detector to be able to apply the new PHOTOM files to all other targets. Ideally, this would be done with fully extended uniform illumination of the detectors, where the flux is known. However, no such target exists in space, and our best calibrators are hot stars with continuum close to a black body. Therefore, to characterise the 2D response of the detectors, a large dataset is required where the star is raster scanned on many locations on the detector. Finally, one downside of using stellar spectra is their relatively low flux farther in the mid-infrared. The lack of signal in channel 4 results in a noisy reference spectrum, or zero flux in these ranges, making the process of calibrating them more difficult.


\section{Conclusions}
\label{sec:conclusions}
In this paper we have shown the importance of a suitable fringe correction for unresolved sources. The pipeline approach (extended fringe flat + residual fringe correction)   removes most of the fringe signal and allows the detection of molecular species. While the method is suitable for most applications, there is a risk of periodic features similar to the expected fringe pattern being weakened or removed by the residual fringe algorithm. Especially for point sources, the PSFF has the potential to out-perform the pipeline approach and provide a higher fidelity correction when required, limited by the reproducibility of the pointing. This approach does not rely on fitting and removing periodic features, making it an inherently safer approach when analysing spectra with highly periodic features.

Intricately linked to the fringes is the spectrophotometric calibration. We demonstrated that a clean on-sky calibration is possible for point sources. This method will be extrapolated to a larger dataset to probe and characterise the 2D response of the detectors in order to create clean files applicable to spatially extended targets observed with the MRS.

Finally, we examined the impact of different fringe correction methods on the detectability of CO, SiO, and OH in the spectra of the K giant HD~37122; and CO in the G dwarf HD~159222. The molecular bands are resolved after effectively removing the fringes. However, we did note a change in the molecular features due to parts of the frequency spectrum being removed by the residual fringe correction. Though this change is subtle and will not impact most science cases, it is something to look out for when running the algorithm. There are several cases where the PSFF does become highly beneficial. Especially for molecules with periodic features (e.g. hydrocarbons), the residual fringe correction may change the shape or remove them altogether. Additionally, the improved spectrophotometric calibration does not require stitching of bands, allowing for better certainty on the continuum, relevant for determining the composition of dust and/or ices. Though its effectiveness is currently limited by the accuracy of the pointing, a library of PSFF fringe flats with small offsets can largely mitigate this issue. The methods presented in this paper have already successfully been applied by \citet{ref:22GrDiTa}, and  show promising results in ongoing molecular studies.

\begin{acknowledgements}
The authors thank Prof. dr. Ewine van Dishoeck for useful comments and suggestions that helped improve the quality of the work presented.\\

Danny Gasman, Ioannis Argyriou, Bart Vandenbussche, and Pierre Royer, thank the European Space Agency (ESA) and the Belgian Federal Science Policy Office (BELSPO) for their support in the framework of the PRODEX Programme. \\
J.A-M. and A.L. acknowledge support by grant PIB2021-127718NB-100 from the Spanish Ministry of Science and Innovation/State Agency of Research MCIN/AEI/10.13039/501100011033 and by “ERDF A way of making Europe”.

P.J.K. acknowledges financial support from the Science Foundation Ireland/Irish Research Council Pathway programme under Grant Number 21/PATH-S/9360.\\

The work presented is the effort of the entire MIRI team and the enthusiasm within the MIRI partnership is a significant factor in its success. MIRI draws on the scientific and technical expertise of the following organisations: Ames Research Center, USA; Airbus Defence and Space, UK; CEA-Irfu, Saclay, France; Centre Spatial de Liége, Belgium; Consejo Superior de Investigaciones Científicas, Spain; Carl Zeiss Optronics, Germany; Chalmers University of Technology, Sweden; Danish Space Research Institute, Denmark; Dublin Institute for Advanced Studies, Ireland; European Space Agency, Netherlands; ETCA, Belgium; ETH Zurich, Switzerland; Goddard Space Flight Center, USA; Institute d'Astrophysique Spatiale, France; Instituto Nacional de Técnica Aeroespacial, Spain; Institute for Astronomy, Edinburgh, UK; Jet Propulsion Laboratory, USA; Laboratoire d'Astrophysique de Marseille (LAM), France; Leiden University, Netherlands; Lockheed Advanced Technology Center (USA); NOVA Opt-IR group at Dwingeloo, Netherlands; Northrop Grumman, USA; Max-Planck Institut für Astronomie (MPIA), Heidelberg, Germany; Laboratoire d’Etudes Spatiales et d'Instrumentation en Astrophysique (LESIA), France; Paul Scherrer Institut, Switzerland; Raytheon Vision Systems, USA; RUAG Aerospace, Switzerland; Rutherford Appleton Laboratory (RAL Space), UK; Space Telescope Science Institute, USA; Toegepast- Natuurwetenschappelijk Onderzoek (TNO-TPD), Netherlands; UK Astronomy Technology Centre, UK; University College London, UK; University of Amsterdam, Netherlands; University of Arizona, USA; University of Bern, Switzerland; University of Cardiff, UK; University of Cologne, Germany; University of Ghent; University of Groningen, Netherlands; University of Leicester, UK; University of Leuven, Belgium; University of Stockholm, Sweden; Utah State University, USA. A portion of this work was carried out at the Jet Propulsion Laboratory, California Institute of Technology, under a contract with the National Aeronautics and Space Administration.\\

We would like to thank the following National and International Funding Agencies for their support of the MIRI development: NASA; ESA; Belgian Science Policy Office; Centre Nationale D'Etudes Spatiales (CNES); Danish National Space Centre; Deutsches Zentrum fur Luft-und Raumfahrt (DLR); Enterprise Ireland; Ministerio De Economiá y Competividad; Netherlands Research School for Astronomy (NOVA); Netherlands Organisation for Scientific Research (NWO); Science and Technology Facilities Council; Swiss Space Office; Swedish National Space Board; UK Space Agency.\\

We take this opportunity to thank the ESA \textit{JWST} Project team and the NASA Goddard ISIM team for their capable technical support in the development of MIRI, its delivery and successful integration.\\

This work has made use of data from the European Space Agency (ESA) mission
{\it Gaia} (\url{https://www.cosmos.esa.int/gaia}), processed by the {\it Gaia}
Data Processing and Analysis Consortium (DPAC,
\url{https://www.cosmos.esa.int/web/gaia/dpac/consortium}). Funding for the DPAC
has been provided by national institutions, in particular the institutions
participating in the {\it Gaia} Multilateral Agreement.\\

This work is based on observations made with the NASA/ESA/CSA James Webb Space Telescope. The data were obtained from the Mikulski Archive for Space Telescopes at the Space Telescope Science Institute, which is operated by the Association of Universities for Research in Astronomy, Inc., under NASA contract NAS 5-03127 for \textit{JWST}; and from the \href{https://jwst.esac.esa.int/archive/}{European \textit{JWST} archive (eJWST)} operated by the ESAC Science Data Centre (ESDC) of the European Space Agency. These observations are associated with program \#1029, 1050.

\end{acknowledgements}

%
%
\bibliographystyle{aa}
\bibliography{bib}

\begin{thebibliography}{49}
\expandafter\ifx\csname natexlab\endcsname\relax\def\natexlab#1{#1}\fi

\bibitem[{{\'{A}}lvarez-Márquez {et~al.}(2023){\'{A}}lvarez-Márquez, Labiano,
  Guillard, Dicken, Argyriou, Patapis, Law, Kavanagh, Larson, Gasman, Mueller,
  Alberts, Brandl, Colina, García-Marín, Jones, Noriega-Crespo, Shivaei,
  Temim, \& Wright}]{ref:22AlLaGu}
{\'{A}}lvarez-Márquez, J., Labiano, A., Guillard, P., {et~al.} 2023, A\&A, in
  press

\bibitem[{{Ardila} {et~al.}(2010){Ardila}, {Van Dyk}, {Makowiecki}, {Stauffer},
  {Song}, {Rho}, {Fajardo-Acosta}, {Hoard}, \& {Wachter}}]{ref:10ArDyMa}
{Ardila}, D.~R., {Van Dyk}, S.~D., {Makowiecki}, W., {et~al.} 2010, \apjs, 191,
  301

\bibitem[{Argyriou(2021)}]{phdthesisYannis}
Argyriou, I. 2021, Calibration of the MIRI instrument on board the James Webb
  Space Telescope,
  \url{https://fys.kuleuven.be/ster/pub/phd-thesis-yannis-argyriou/PhD_Thesis_IoannisArgyriou.pdf}

\bibitem[{{Argyriou} {et~al.}(2020{\natexlab{a}}){Argyriou}, {Rieke},
  {Ressler}, {G{\'a}sp{\'a}r}, \& {Vandenbussche}}]{ref:20ArRiRe}
{Argyriou}, I., {Rieke}, G.~H., {Ressler}, M.~E., {G{\'a}sp{\'a}r}, A., \&
  {Vandenbussche}, B. 2020{\natexlab{a}}, in Society of Photo-Optical
  Instrumentation Engineers (SPIE) Conference Series, Vol. 11454, Society of
  Photo-Optical Instrumentation Engineers (SPIE) Conference Series, 114541P

\bibitem[{{Argyriou} {et~al.}(2020{\natexlab{b}}){Argyriou}, Wells, Glasse,
  Lee, Royer, Vandenbussche, Malumuth, Glauser, Kavanagh, Labiano, Lahuis,
  Mueller, \& Patapis}]{ref:20ArWeGl}
{Argyriou}, I., Wells, M., Glasse, A., {et~al.} 2020{\natexlab{b}}, A\&A, 640,
  A150

\bibitem[{{Aringer} {et~al.}(2016){Aringer}, {Girardi}, {Nowotny}, {Marigo}, \&
  {Bressan}}]{ref:16ArGiNo}
{Aringer}, B., {Girardi}, L., {Nowotny}, W., {Marigo}, P., \& {Bressan}, A.
  2016, \mnras, 457, 3611

\bibitem[{{Astropy Collaboration} {et~al.}(2018){Astropy Collaboration},
  {Price-Whelan}, {Sip{\H{o}}cz}, {G{\"u}nther}, {Lim}, {Crawford}, {Conseil},
  {Shupe}, {Craig}, {Dencheva}, {Ginsburg}, {Vand erPlas}, {Bradley},
  {P{\'e}rez-Su{\'a}rez}, {de Val-Borro}, {Aldcroft}, {Cruz}, {Robitaille},
  {Tollerud}, {Ardelean}, {Babej}, {Bach}, {Bachetti}, {Bakanov}, {Bamford},
  {Barentsen}, {Barmby}, {Baumbach}, {Berry}, {Biscani}, {Boquien}, {Bostroem},
  {Bouma}, {Brammer}, {Bray}, {Breytenbach}, {Buddelmeijer}, {Burke},
  {Calderone}, {Cano Rodr{\'\i}guez}, {Cara}, {Cardoso}, {Cheedella}, {Copin},
  {Corrales}, {Crichton}, {D'Avella}, {Deil}, {Depagne}, {Dietrich}, {Donath},
  {Droettboom}, {Earl}, {Erben}, {Fabbro}, {Ferreira}, {Finethy}, {Fox},
  {Garrison}, {Gibbons}, {Goldstein}, {Gommers}, {Greco}, {Greenfield},
  {Groener}, {Grollier}, {Hagen}, {Hirst}, {Homeier}, {Horton}, {Hosseinzadeh},
  {Hu}, {Hunkeler}, {Ivezi{\'c}}, {Jain}, {Jenness}, {Kanarek}, {Kendrew},
  {Kern}, {Kerzendorf}, {Khvalko}, {King}, {Kirkby}, {Kulkarni}, {Kumar},
  {Lee}, {Lenz}, {Littlefair}, {Ma}, {Macleod}, {Mastropietro}, {McCully},
  {Montagnac}, {Morris}, {Mueller}, {Mumford}, {Muna}, {Murphy}, {Nelson},
  {Nguyen}, {Ninan}, {N{\"o}the}, {Ogaz}, {Oh}, {Parejko}, {Parley}, {Pascual},
  {Patil}, {Patil}, {Plunkett}, {Prochaska}, {Rastogi}, {Reddy Janga},
  {Sabater}, {Sakurikar}, {Seifert}, {Sherbert}, {Sherwood-Taylor}, {Shih},
  {Sick}, {Silbiger}, {Singanamalla}, {Singer}, {Sladen}, {Sooley},
  {Sornarajah}, {Streicher}, {Teuben}, {Thomas}, {Tremblay}, {Turner},
  {Terr{\'o}n}, {van Kerkwijk}, {de la Vega}, {Watkins}, {Weaver}, {Whitmore},
  {Woillez}, {Zabalza}, \& {Astropy Contributors}}]{astropy:2018}
{Astropy Collaboration}, {Price-Whelan}, A.~M., {Sip{\H{o}}cz}, B.~M., {et~al.}
  2018, \aj, 156, 123

\bibitem[{{Astropy Collaboration} {et~al.}(2013){Astropy Collaboration},
  {Robitaille}, {Tollerud}, {Greenfield}, {Droettboom}, {Bray}, {Aldcroft},
  {Davis}, {Ginsburg}, {Price-Whelan}, {Kerzendorf}, {Conley}, {Crighton},
  {Barbary}, {Muna}, {Ferguson}, {Grollier}, {Parikh}, {Nair}, {Unther},
  {Deil}, {Woillez}, {Conseil}, {Kramer}, {Turner}, {Singer}, {Fox}, {Weaver},
  {Zabalza}, {Edwards}, {Azalee Bostroem}, {Burke}, {Casey}, {Crawford},
  {Dencheva}, {Ely}, {Jenness}, {Labrie}, {Lim}, {Pierfederici}, {Pontzen},
  {Ptak}, {Refsdal}, {Servillat}, \& {Streicher}}]{astropy:2013}
{Astropy Collaboration}, {Robitaille}, T.~P., {Tollerud}, E.~J., {et~al.} 2013,
  \aap, 558, A33

\bibitem[{{Barton} {et~al.}(2013){Barton}, {Yurchenko}, \&
  {Tennyson}}]{ref:13BaYuTe}
{Barton}, E.~J., {Yurchenko}, S.~N., \& {Tennyson}, J. 2013, \mnras, 434, 1469

\bibitem[{{Bohlin} {et~al.}(2020){Bohlin}, {Hubeny}, \& {Rauch}}]{ref:20BoHuRa}
{Bohlin}, R.~C., {Hubeny}, I., \& {Rauch}, T. 2020, \aj, 160, 21

\bibitem[{{Bohlin} {et~al.}(2017){Bohlin}, {M{\'e}sz{\'a}ros}, {Fleming},
  {Gordon}, {Koekemoer}, \& {Kov{\'a}cs}}]{ref:17BoMeFl}
{Bohlin}, R.~C., {M{\'e}sz{\'a}ros}, S., {Fleming}, S.~W., {et~al.} 2017, \aj,
  153, 234

\bibitem[{{Carnall}(2017)}]{ref:17Ca}
{Carnall}, A.~C. 2017, arXiv e-prints, arXiv:1705.05165

\bibitem[{{Cohen} {et~al.}(2003){Cohen}, {Megeath}, {Hammersley},
  {Mart{\'\i}n-Luis}, \& {Stauffer}}]{ref:03CoMeHa}
{Cohen}, M., {Megeath}, S.~T., {Hammersley}, P.~L., {Mart{\'\i}n-Luis}, F., \&
  {Stauffer}, J. 2003, \aj, 125, 2645

\bibitem[{{Decin} {et~al.}(2004){Decin}, {Morris}, {Appleton}, {Charmandaris},
  {Armus}, \& {Houck}}]{ref:04DeMoAp}
{Decin}, L., {Morris}, P.~W., {Appleton}, P.~N., {et~al.} 2004, \apjs, 154, 408

\bibitem[{Decin {et~al.}(2003)Decin, Vandenbussche, Waelkens, Eriksson,
  Gustafsson, Plez, \& Sauval}]{ref:03DeVaWa}
Decin, L., Vandenbussche, B., Waelkens, C., {et~al.} 2003, A\&A, 400

\bibitem[{{Engelbracht} {et~al.}(2007){Engelbracht}, {Blaylock}, {Su}, {Rho},
  {Rieke}, {Muzerolle}, {Padgett}, {Hines}, {Gordon}, {Fadda},
  {Noriega-Crespo}, {Kelly}, {Latter}, {Hinz}, {Misselt}, {Morrison},
  {Stansberry}, {Shupe}, {Stolovy}, {Wheaton}, {Young}, {Neugebauer},
  {Wachter}, {P{\'e}rez-Gonz{\'a}lez}, {Frayer}, \& {Marleau}}]{ref:07EnBlSu}
{Engelbracht}, C.~W., {Blaylock}, M., {Su}, K.~Y.~L., {et~al.} 2007, \pasp,
  119, 994

\bibitem[{{Fabricius} {et~al.}(2021){Fabricius}, {Luri}, {Arenou}, {Babusiaux},
  {Helmi}, {Muraveva}, {Reyl{\'e}}, {Spoto}, {Vallenari}, {Antoja}, {Balbinot},
  {Barache}, {Bauchet}, {Bragaglia}, {Busonero}, {Cantat-Gaudin}, {Carrasco},
  {Diakit{\'e}}, {Fabrizio}, {Figueras}, {Garcia-Gutierrez}, {Garofalo},
  {Jordi}, {Kervella}, {Khanna}, {Leclerc}, {Licata}, {Lambert}, {Marrese},
  {Masip}, {Ramos}, {Robichon}, {Robin}, {Romero-G{\'o}mez}, {Rubele}, \&
  {Weiler}}]{ref:21gaia2}
{Fabricius}, C., {Luri}, X., {Arenou}, F., {et~al.} 2021, \aap, 649, A5

\bibitem[{{Gaia Collaboration} {et~al.}(2021){Gaia Collaboration}, {Brown},
  {Vallenari}, {Prusti}, {de Bruijne}, {Babusiaux}, {Biermann}, {Creevey},
  {Evans}, {Eyer}, {Hutton}, {Jansen}, {Jordi}, {Klioner}, {Lammers},
  {Lindegren}, {Luri}, {Mignard}, {Panem}, {Pourbaix}, {Randich}, {Sartoretti},
  {Soubiran}, {Walton}, {Arenou}, {Bailer-Jones}, {Bastian}, {Cropper},
  {Drimmel}, {Katz}, {Lattanzi}, {van Leeuwen}, {Bakker}, {Cacciari},
  {Casta{\~n}eda}, {De Angeli}, {Ducourant}, {Fabricius}, {Fouesneau},
  {Fr{\'e}mat}, {Guerra}, {Guerrier}, {Guiraud}, {Jean-Antoine Piccolo},
  {Masana}, {Messineo}, {Mowlavi}, {Nicolas}, {Nienartowicz}, {Pailler},
  {Panuzzo}, {Riclet}, {Roux}, {Seabroke}, {Sordo}, {Tanga}, {Th{\'e}venin},
  {Gracia-Abril}, {Portell}, {Teyssier}, {Altmann}, {Andrae}, {Bellas-Velidis},
  {Benson}, {Berthier}, {Blomme}, {Brugaletta}, {Burgess}, {Busso}, {Carry},
  {Cellino}, {Cheek}, {Clementini}, {Damerdji}, {Davidson}, {Delchambre},
  {Dell'Oro}, {Fern{\'a}ndez-Hern{\'a}ndez}, {Galluccio}, {Garc{\'\i}a-Lario},
  {Garcia-Reinaldos}, {Gonz{\'a}lez-N{\'u}{\~n}ez}, {Gosset}, {Haigron},
  {Halbwachs}, {Hambly}, {Harrison}, {Hatzidimitriou}, {Heiter},
  {Hern{\'a}ndez}, {Hestroffer}, {Hodgkin}, {Holl}, {Jan{\ss}en}, {Jevardat de
  Fombelle}, {Jordan}, {Krone-Martins}, {Lanzafame}, {L{\"o}ffler}, {Lorca},
  {Manteiga}, {Marchal}, {Marrese}, {Moitinho}, {Mora}, {Muinonen}, {Osborne},
  {Pancino}, {Pauwels}, {Petit}, {Recio-Blanco}, {Richards}, {Riello},
  {Rimoldini}, {Robin}, {Roegiers}, {Rybizki}, {Sarro}, {Siopis}, {Smith},
  {Sozzetti}, {Ulla}, {Utrilla}, {van Leeuwen}, {van Reeven}, {Abbas}, {Abreu
  Aramburu}, {Accart}, {Aerts}, {Aguado}, {Ajaj}, {Altavilla}, {{\'A}lvarez},
  {{\'A}lvarez Cid-Fuentes}, {Alves}, {Anderson}, {Anglada Varela}, {Antoja},
  {Audard}, {Baines}, {Baker}, {Balaguer-N{\'u}{\~n}ez}, {Balbinot}, {Balog},
  {Barache}, {Barbato}, {Barros}, {Barstow}, {Bartolom{\'e}}, {Bassilana},
  {Bauchet}, {Baudesson-Stella}, {Becciani}, {Bellazzini}, {Bernet}, {Bertone},
  {Bianchi}, {Blanco-Cuaresma}, {Boch}, {Bombrun}, {Bossini}, {Bouquillon},
  {Bragaglia}, {Bramante}, {Breedt}, {Bressan}, {Brouillet}, {Bucciarelli},
  {Burlacu}, {Busonero}, {Butkevich}, {Buzzi}, {Caffau}, {Cancelliere},
  {C{\'a}novas}, {Cantat-Gaudin}, {Carballo}, {Carlucci}, {Carnerero},
  {Carrasco}, {Casamiquela}, {Castellani}, {Castro-Ginard}, {Castro Sampol},
  {Chaoul}, {Charlot}, {Chemin}, {Chiavassa}, {Cioni}, {Comoretto}, {Cooper},
  {Cornez}, {Cowell}, {Crifo}, {Crosta}, {Crowley}, {Dafonte}, {Dapergolas},
  {David}, {David}, {de Laverny}, {De Luise}, {De March}, {De Ridder}, {de
  Souza}, {de Teodoro}, {de Torres}, {del Peloso}, {del Pozo}, {Delbo},
  {Delgado}, {Delgado}, {Delisle}, {Di Matteo}, {Diakite}, {Diener},
  {Distefano}, {Dolding}, {Eappachen}, {Edvardsson}, {Enke}, {Esquej}, {Fabre},
  {Fabrizio}, {Faigler}, {Fedorets}, {Fernique}, {Fienga}, {Figueras},
  {Fouron}, {Fragkoudi}, {Fraile}, {Franke}, {Gai}, {Garabato},
  {Garcia-Gutierrez}, {Garc{\'\i}a-Torres}, {Garofalo}, {Gavras}, {Gerlach},
  {Geyer}, {Giacobbe}, {Gilmore}, {Girona}, {Giuffrida}, {Gomel}, {Gomez},
  {Gonzalez-Santamaria}, {Gonz{\'a}lez-Vidal}, {Granvik},
  {Guti{\'e}rrez-S{\'a}nchez}, {Guy}, {Hauser}, {Haywood}, {Helmi}, {Hidalgo},
  {Hilger}, {H{\l}adczuk}, {Hobbs}, {Holland}, {Huckle}, {Jasniewicz},
  {Jonker}, {Juaristi Campillo}, {Julbe}, {Karbevska}, {Kervella}, {Khanna},
  {Kochoska}, {Kontizas}, {Kordopatis}, {Korn}, {Kostrzewa-Rutkowska},
  {Kruszy{\'n}ska}, {Lambert}, {Lanza}, {Lasne}, {Le Campion}, {Le Fustec},
  {Lebreton}, {Lebzelter}, {Leccia}, {Leclerc}, {Lecoeur-Taibi}, {Liao},
  {Licata}, {Lindstr{\o}m}, {Lister}, {Livanou}, {Lobel}, {Madrero Pardo},
  {Managau}, {Mann}, {Marchant}, {Marconi}, {Marcos Santos}, {Marinoni},
  {Marocco}, {Marshall}, {Martin Polo}, {Mart{\'\i}n-Fleitas}, {Masip},
  {Massari}, {Mastrobuono-Battisti}, {Mazeh}, {McMillan}, {Messina},
  {Michalik}, {Millar}, {Mints}, {Molina}, {Molinaro}, {Moln{\'a}r},
  {Montegriffo}, {Mor}, {Morbidelli}, {Morel}, {Morris}, {Mulone}, {Munoz},
  {Muraveva}, {Murphy}, {Musella}, {Noval}, {Ord{\'e}novic}, {Orr{\`u}},
  {Osinde}, {Pagani}, {Pagano}, {Palaversa}, {Palicio}, {Panahi}, {Pawlak},
  {Pe{\~n}alosa Esteller}, {Penttil{\"a}}, {Piersimoni}, {Pineau}, {Plachy},
  {Plum}, {Poggio}, {Poretti}, {Poujoulet}, {Pr{\v{s}}a}, {Pulone}, {Racero},
  {Ragaini}, {Rainer}, {Raiteri}, {Rambaux}, {Ramos}, {Ramos-Lerate}, {Re
  Fiorentin}, {Regibo}, {Reyl{\'e}}, {Ripepi}, {Riva}, {Rixon}, {Robichon},
  {Robin}, {Roelens}, {Rohrbasser}, {Romero-G{\'o}mez}, {Rowell}, {Royer},
  {Rybicki}, {Sadowski}, {Sagrist{\`a} Sell{\'e}s}, {Sahlmann}, {Salgado},
  {Salguero}, {Samaras}, {Sanchez Gimenez}, {Sanna}, {Santove{\~n}a},
  {Sarasso}, {Schultheis}, {Sciacca}, {Segol}, {Segovia}, {S{\'e}gransan},
  {Semeux}, {Shahaf}, {Siddiqui}, {Siebert}, {Siltala}, {Slezak}, {Smart},
  {Solano}, {Solitro}, {Souami}, {Souchay}, {Spagna}, {Spoto}, {Steele},
  {Steidelm{\"u}ller}, {Stephenson}, {S{\"u}veges}, {Szabados}, {Szegedi-Elek},
  {Taris}, {Tauran}, {Taylor}, {Teixeira}, {Thuillot}, {Tonello}, {Torra},
  {Torra}, {Turon}, {Unger}, {Vaillant}, {van Dillen}, {Vanel}, {Vecchiato},
  {Viala}, {Vicente}, {Voutsinas}, {Weiler}, {Wevers}, {Wyrzykowski}, {Yoldas},
  {Yvard}, {Zhao}, {Zorec}, {Zucker}, {Zurbach}, \& {Zwitter}}]{ref:21gaia}
{Gaia Collaboration}, {Brown}, A.~G.~A., {Vallenari}, A., {et~al.} 2021, \aap,
  649, A1

\bibitem[{{Gaia Collaboration} {et~al.}(2016){Gaia Collaboration}, Prusti,
  de~Bruijne, Brown, Vallenari, Babusiaux, Bailer-Jones, Bastian, Biermann,
  Evans, Eyer, Jansen, Jordi, Klioner, Lammers, Lindegren, Luri, Mignard,
  Milligan, Panem, Poinsignon, Pourbaix, Randich, Sarri, Sartoretti, Siddiqui,
  Soubiran, Valette, van Leeuwen, Walton, Aerts, Arenou, Cropper, Drimmel,
  H{\o}g, Katz, Lattanzi, O'Mullane, Grebel, Holland, Huc, Passot, Bramante,
  Cacciari, Casta{\~{n} }eda, Chaoul, Cheek, Angeli, Fabricius, Guerra,
  Hern{\'{a}}ndez, Jean-Antoine-Piccolo, Masana, Messineo, Mowlavi,
  Nienartowicz, Ord{\'{o}}{\~{n}}ez-Blanco, Panuzzo, Portell, Richards, Riello,
  Seabroke, Tanga, Th{\'{e}}venin, Torra, Els, Gracia-Abril, Comoretto,
  Garcia-Reinaldos, Lock, Mercier, Altmann, Andrae, Astraatmadja,
  Bellas-Velidis, Benson, Berthier, Blomme, Busso, Carry, Cellino, Clementini,
  Cowell, Creevey, Cuypers, Davidson, Ridder, de~Torres, Delchambre, Dell'Oro,
  Ducourant, Fr{\'{e}}mat, Garc{\'{\i}}a-Torres, Gosset, Halbwachs, Hambly,
  Harrison, Hauser, Hestroffer, Hodgkin, Huckle, Hutton, Jasniewicz, Jordan,
  Kontizas, Korn, Lanzafame, Manteiga, Moitinho, Muinonen, Osinde, Pancino,
  Pauwels, Petit, Recio-Blanco, Robin, Sarro, Siopis, Smith, Smith, Sozzetti,
  Thuillot, van Reeven, Viala, Abbas, Aramburu, Accart, Aguado, Allan, Allasia,
  Altavilla, {\'{A}}lvarez, Alves, Anderson, Andrei, Varela, Antiche, Antoja,
  Ant{\'{o}}n, Arcay, Atzei, Ayache, Bach, Baker, Balaguer-N{\'{u}}{\~{n}}ez,
  Barache, Barata, Barbier, Barblan, Baroni, y~Navascu{\'{e}}s, Barros,
  Barstow, Becciani, Bellazzini, Bellei, Garc{\'{\i}}a, Belokurov, Bendjoya,
  Berihuete, Bianchi, Bienaym{\'{e}}, Billebaud, Blagorodnova, Blanco-Cuaresma,
  Boch, Bombrun, Borrachero, Bouquillon, Bourda, Bouy, Bragaglia, Breddels,
  Brouillet, Brüsemeister, Bucciarelli, Budnik, Burgess, Burgon, Burlacu,
  Busonero, Buzzi, Caffau, Cambras, Campbell, Cancelliere, Cantat-Gaudin,
  Carlucci, Carrasco, Castellani, Charlot, Charnas, Charvet, Chassat,
  Chiavassa, Clotet, Cocozza, Collins, Collins, Costigan, Crifo, Cross, Crosta,
  Crowley, Dafonte, Damerdji, Dapergolas, David, David, Cat, de~Felice,
  de~Laverny, Luise, March, de~Martino, de~Souza, Debosscher, del Pozo, Delbo,
  Delgado, Delgado, di~Marco, Matteo, Diakite, Distefano, Dolding, Anjos,
  Drazinos, Dur{\'{a}}n, Dzigan, Ecale, Edvardsson, Enke, Erdmann, Escolar,
  Espina, Evans, Bontemps, Fabre, Fabrizio, Faigler, Falc{\~{a}}o, Casas, Faye,
  Federici, Fedorets, Fern{\'{a}}ndez-Hern{\'{a}}ndez, Fernique, Fienga,
  Figueras, Filippi, Findeisen, Fonti, Fouesneau, Fraile, Fraser, Fuchs,
  Furnell, Gai, Galleti, Galluccio, Garabato, Garc{\'{\i}}a-Sedano, Gar{\'{e}},
  Garofalo, Garralda, Gavras, Gerssen, Geyer, Gilmore, Girona, Giuffrida,
  Gomes, Gonz{\'{a}}lez-Marcos, Gonz{\'{a}}lez-N{\'{u}}{\~{n}}ez,
  Gonz{\'{a}}lez-Vidal, Granvik, Guerrier, Guillout, Guiraud, G{\'{u}}rpide,
  Guti{\'{e}}rrez-S{\'{a}}nchez, Guy, Haigron, Hatzidimitriou, Haywood, Heiter,
  Helmi, Hobbs, Hofmann, Holl, Holland, Hunt, Hypki, Icardi, Irwin,
  de~Fombelle, Jofr{\'{e}}, Jonker, Jorissen, Julbe, Karampelas, Kochoska,
  Kohley, Kolenberg, Kontizas, Koposov, Kordopatis, Koubsky, Kowalczyk,
  Krone-Martins, Kudryashova, Kull, Bachchan, Lacoste-Seris, Lanza, Lavigne,
  Poncin-Lafitte, Lebreton, Lebzelter, Leccia, Leclerc, Lecoeur-Taibi,
  Lemaitre, Lenhardt, Leroux, Liao, Licata, Lindstr{\o}m, Lister, Livanou,
  Lobel, Löffler, L{\'{o}}pez, Lopez-Lozano, Lorenz, Loureiro, MacDonald,
  Fernandes, Managau, Mann, Mantelet, Marchal, Marchant, Marconi, Marie,
  Marinoni, Marrese, Marschalk{\'{o}}, Marshall, Mart{\'{\i}}n-Fleitas,
  Martino, Mary, Matijevi{\v{c}}, Mazeh, McMillan, Messina, Mestre, Michalik,
  Millar, Miranda, Molina, Molinaro, Molinaro, Moln{\'{a}}r, Moniez,
  Montegriffo, Monteiro, Mor, Mora, Morbidelli, Morel, Morgenthaler, Morley,
  Morris, Mulone, Muraveva, Musella, Narbonne, Nelemans, Nicastro, Noval,
  Ord{\'{e}}novic, Ordieres-Mer{\'{e}}, Osborne, Pagani, Pagano, Pailler,
  Palacin, Palaversa, Parsons, Paulsen, Pecoraro, Pedrosa, Pentikäinen,
  Pereira, Pichon, Piersimoni, Pineau, Plachy, Plum, Poujoulet, Pr{\v{s}}a,
  Pulone, Ragaini, Rago, Rambaux, Ramos-Lerate, Ranalli, Rauw, Read, Regibo,
  Renk, Reyl{\'{e}}, Ribeiro, Rimoldini, Ripepi, Riva, Rixon, Roelens,
  Romero-G{\'{o}}mez, Rowell, Royer, Rudolph, Ruiz-Dern, Sadowski,
  Sell{\'{e}}s, Sahlmann, Salgado, Salguero, Sarasso, Savietto, Schnorhk,
  Schultheis, Sciacca, Segol, Segovia, Segransan, Serpell, Shih, Smareglia,
  Smart, Smith, Solano, Solitro, Sordo, Nieto, Souchay, Spagna, Spoto, Stampa,
  Steele, Steidelmüller, Stephenson, Stoev, Suess, Süveges, Surdej, Szabados,
  Szegedi-Elek, Tapiador, Taris, Tauran, Taylor, Teixeira, Terrett, Tingley,
  Trager, Turon, Ulla, Utrilla, Valentini, van Elteren, Hemelryck, van Leeuwen,
  Varadi, Vecchiato, Veljanoski, Via, Vicente, Vogt, Voss, Votruba, Voutsinas,
  Walmsley, Weiler, Weingrill, Werner, Wevers, Whitehead, Wyrzykowski, Yoldas,
  {\v{Z}}erjal, Zucker, Zurbach, Zwitter, Alecu, Allen, Prieto, Amorim,
  Anglada-Escud{\'{e}}, Arsenijevic, Azaz, Balm, Beck, Bernstein, Bigot,
  Bijaoui, Blasco, Bonfigli, Bono, Boudreault, Bressan, Brown, Brunet,
  Bunclark, Buonanno, Butkevich, Carret, Carrion, Chemin, Ch{\'{e}}reau,
  Corcione, Darmigny, de~Boer, de~Teodoro, de~Zeeuw, Luche, Domingues, Dubath,
  Fodor, Fr{\'{e}}zouls, Fries, Fustes, Fyfe, Gallardo, Gallegos, Gardiol,
  Gebran, Gomboc, G{\'{o}}mez, Grux, Gueguen, Heyrovsky, Hoar, Iannicola,
  Parache, Janotto, Joliet, Jonckheere, Keil, Kim, Klagyivik, Klar, Knude,
  Kochukhov, Kolka, Kos, Kutka, Lainey, LeBouquin, Liu, Loreggia, Makarov,
  Marseille, Martayan, Martinez-Rubi, Massart, Meynadier, Mignot, Munari,
  Nguyen, Nordlander, Ocvirk, O'Flaherty, Sanz, Ortiz, Osorio, Oszkiewicz,
  Ouzounis, Palmer, Park, Pasquato, Peltzer, Peralta, P{\'{e}}turaud,
  Pieniluoma, Pigozzi, Poels, Prat, Prod'homme, Raison, Rebordao, Risquez,
  Rocca-Volmerange, Rosen, Ruiz-Fuertes, Russo, Sembay, Vizcaino, Short,
  Siebert, Silva, Sinachopoulos, Slezak, Soffel, Sosnowska, Strai{\v{z}}ys, ter
  Linden, Terrell, Theil, Tiede, Troisi, Tsalmantza, Tur, Vaccari, Vachier,
  Valles, Hamme, Veltz, Virtanen, Wallut, Wichmann, Wilkinson, Ziaeepour, \&
  Zschocke}]{ref:16gaia}
{Gaia Collaboration}, Prusti, T., de~Bruijne, J. H.~J., {et~al.} 2016, A\&A,
  595, A1

\bibitem[{{Garc{\'\i}a-Bernete} {et~al.}(2022){Garc{\'\i}a-Bernete},
  {Rigopoulou}, {Alonso-Herrero}, {Donnan}, {Roche}, {Pereira-Santaella},
  {Labiano}, {Peralta de Arriba}, {Izumi}, {Ramos Almeida}, {Shimizu},
  {H{\"o}nig}, {Garc{\'\i}a-Burillo}, {Rosario}, {Ward}, {Bellocchi}, {Hicks},
  {Fuller}, \& {Packham}}]{ref:22GaRiAl}
{Garc{\'\i}a-Bernete}, I., {Rigopoulou}, D., {Alonso-Herrero}, A., {et~al.}
  2022, \aap, 666, L5

\bibitem[{{Gardner} {et~al.}(2006){Gardner}, {Mather}, {Clampin}, {Doyon},
  {Greenhouse}, {Hammel}, {Hutchings}, {Jakobsen}, {Lilly}, {Long}, {Lunine},
  {McCaughrean}, {Mountain}, {Nella}, {Rieke}, {Rieke}, {Rix}, {Smith},
  {Sonneborn}, {Stiavelli}, {Stockman}, {Windhorst}, \&
  {Wright}}]{ref:06GaMaCl}
{Gardner}, J.~P., {Mather}, J.~C., {Clampin}, M., {et~al.} 2006, \ssr, 123, 485

\bibitem[{Gordon {et~al.}(2022)Gordon, Bohlin, Sloan, Rieke, Volk, Boyer,
  Muzerolle, Schlawin, Deustua, Hines, Kraemer, Mullally, \& Su}]{ref:22GoBoSl}
Gordon, K.~D., Bohlin, R., Sloan, G.~C., {et~al.} 2022, AJ, 163, 13

\bibitem[{Grant {et~al.}(2023)Grant, van Dishoeck, Tabone, Gasman, Henning,
  Kamp, Güdel, Lagage, Bettoni, Perotti, Christiaens, Samland, Arabhavi,
  Argyriou, Abergel, Absil, Barrado, Boccaletti, Bouwman, Garatti, Geers,
  Glauser, Guadarrama, Jang, Kanwar, Lahuis, Morales-Calderón, Mueller,
  Nehmé, Olofsson, Pantin, Pawellek, Ray, Rodgers-Lee, Scheithauer, Schreiber,
  Schwarz, Temmink, Vandenbussche, Vlasblom, Waters, Wright, Colina, Greve,
  Justannont, \& Östlin}]{ref:22GrDiTa}
Grant, S.~L., van Dishoeck, E.~F., Tabone, B., {et~al.} 2023, A\&A, accepted

\bibitem[{{Gray} {et~al.}(2003){Gray}, {Corbally}, {Garrison}, {McFadden}, \&
  {Robinson}}]{ref:03GrCoGa}
{Gray}, R.~O., {Corbally}, C.~J., {Garrison}, R.~F., {McFadden}, M.~T., \&
  {Robinson}, P.~E. 2003, \aj, 126, 2048

\bibitem[{{Heras} {et~al.}(2002){Heras}, {Shipman}, {Price}, {de Graauw},
  {Walker}, {Jourdain de Muizon}, {Kessler}, {Prusti}, {Decin},
  {Vandenbussche}, \& {Waters}}]{ref:02HeShPr}
{Heras}, A.~M., {Shipman}, R.~F., {Price}, S.~D., {et~al.} 2002, \aap, 394, 539

\bibitem[{{Houck} {et~al.}(2004){Houck}, {Roellig}, {van Cleve}, {Forrest},
  {Herter}, {Lawrence}, {Matthews}, {Reitsema}, {Soifer}, {Watson}, {Weedman},
  {Huisjen}, {Troeltzsch}, {Barry}, {Bernard-Salas}, {Blacken}, {Brandl},
  {Charmandaris}, {Devost}, {Gull}, {Hall}, {Henderson}, {Higdon}, {Pirger},
  {Schoenwald}, {Sloan}, {Uchida}, {Appleton}, {Armus}, {Burgdorf},
  {Fajardo-Acosta}, {Grillmair}, {Ingalls}, {Morris}, \&
  {Teplitz}}]{ref:04HoRoCl}
{Houck}, J.~R., {Roellig}, T.~L., {van Cleve}, J., {et~al.} 2004, \apjs, 154,
  18

\bibitem[{Kester {et~al.}(2003)Kester, Beintema, \& Lutz}]{ref:03KeBeLu}
Kester, D. J.~M., Beintema, D.~A., \& Lutz, D. 2003, in {The Calibration Legacy
  of the ISO Mission}, ed. L.~Metcalfe, A.~Salama, S.~B. Peschke, \& M.~F.
  Kessler, Vol. 481, 375

\bibitem[{{Labiano} {et~al.}(2021){Labiano}, {Argyriou},
  {{\'A}lvarez-M{\'a}rquez}, {Glasse}, {Glauser}, {Patapis}, {Law}, {Brandl},
  {Justtanont}, {Lahuis}, {Mart{\'\i}nez-Galarza}, {Mueller}, {Noriega-Crespo},
  {Royer}, {Shaughnessy}, \& {Vandenbussche}}]{ref:21LaArAl}
{Labiano}, A., {Argyriou}, I., {{\'A}lvarez-M{\'a}rquez}, J., {et~al.} 2021,
  \aap, 656, A57

\bibitem[{Labiano {et~al.}(2016)Labiano, Azzollini, Bailey, Beard, Dicken,
  García-Marín, Geers, Glasse, Glauser, Gordon, Justtanont, Klaassen, Lahuis,
  Law, Morrison, Müller, Rieke, B.Vandenbussche, \& Wright}]{ref:16LaAzBa}
Labiano, A., Azzollini, R., Bailey, J., {et~al.} 2016, in Observatory
  Operations: Strategies, Processes, and Systems VI, Vol. 9910, {Proc.\ The
  MIRI Medium Resolution Spectrometer Calibration Pipeline}

\bibitem[{Lahuis \& Boogert(2003)}]{ref:03LaBo}
Lahuis, F. \& Boogert, A. 2003, in {}, ed. C.~L. Curry \& M.~Fich, 335

\bibitem[{Lahuis \& van Dishoeck(2000)}]{ref:00LaDi}
Lahuis, F. \& van Dishoeck, E.~F. 2000, A\&A, 355, 699

\bibitem[{{Lau} {et~al.}(2022){Lau}, {Hankins}, {Han}, {Argyriou}, {Corcoran},
  {Eldridge}, {Endo}, {Fox}, {Garcia Marin}, {Gull}, {Jones}, {Hamaguchi},
  {Lamberts}, {Law}, {Madura}, {Marchenko}, {Matsuhara}, {Moffat}, {Morris},
  {Morris}, {Onaka}, {Ressler}, {Richardson}, {Russell}, {Sanchez-Bermudez},
  {Smith}, {Soulain}, {Stevens}, {Tuthill}, {Weigelt}, {Williams}, \&
  {Yamaguchi}}]{ref:22Lau}
{Lau}, R.~M., {Hankins}, M.~J., {Han}, Y., {et~al.} 2022, Nature Astronomy, 6,
  1308

\bibitem[{{Lebouteiller} {et~al.}(2011){Lebouteiller}, {Barry}, {Spoon},
  {Bernard-Salas}, {Sloan}, {Houck}, \& {Weedman}}]{ref:11LeBaSp}
{Lebouteiller}, V., {Barry}, D.~J., {Spoon}, H.~W.~W., {et~al.} 2011, \apjs,
  196, 8

\bibitem[{{Li} {et~al.}(2015){Li}, {Gordon}, {Rothman}, {Tan}, {Hu}, {Kassi},
  {Campargue}, \& {Medvedev}}]{ref:15LiGoRo}
{Li}, G., {Gordon}, I.~E., {Rothman}, L.~S., {et~al.} 2015, \apjs, 216, 15

\bibitem[{{Mahdi} {et~al.}(2016){Mahdi}, {Soubiran}, {Blanco-Cuaresma}, \&
  {Chemin}}]{ref:16MaBlCh}
{Mahdi}, D., {Soubiran}, C., {Blanco-Cuaresma}, S., \& {Chemin}, L. 2016, \aap,
  587, A131

\bibitem[{Malumuth {et~al.}(2003)Malumuth, Hill, Gull, Woodgate, Bowers,
  Kimble, Lindler, Plait, \& Blouke}]{ref:03MaHiGu}
Malumuth, E., Hill, R., Gull, T., {et~al.} 2003, Publications of the
  Astronomical Society of the Pacific, 115, 218

\bibitem[{{Meixner} {et~al.}(2006){Meixner}, {Gordon}, {Indebetouw}, {Hora},
  {Whitney}, {Blum}, {Reach}, {Bernard}, {Meade}, {Babler}, {Engelbracht},
  {For}, {Misselt}, {Vijh}, {Leitherer}, {Cohen}, {Churchwell}, {Boulanger},
  {Frogel}, {Fukui}, {Gallagher}, {Gorjian}, {Harris}, {Kelly}, {Kawamura},
  {Kim}, {Latter}, {Madden}, {Markwick-Kemper}, {Mizuno}, {Mizuno}, {Mould},
  {Nota}, {Oey}, {Olsen}, {Onishi}, {Paladini}, {Panagia}, {Perez-Gonzalez},
  {Shibai}, {Sato}, {Smith}, {Staveley-Smith}, {Tielens}, {Ueta}, {van Dyk},
  {Volk}, {Werner}, \& {Zaritsky}}]{ref:06MeGoIn}
{Meixner}, M., {Gordon}, K.~D., {Indebetouw}, R., {et~al.} 2006, \aj, 132, 2268

\bibitem[{Miles {et~al.}(2022)Miles, Biller, Patapis, Worthen, Rickman, Hoch,
  Skemer, Perrin, Chen, Mukherjee, Morley, Moran, Bonnefoy, Petrus, Carter,
  Choquet, Hinkley, Ward-Duong, Leisenring, Millar-Blanchaer, Pueyo, Ray,
  Stapelfeldt, Stone, Wang, Absil, Balmer, Boccaletti, Bonavita, Booth, Bowler,
  Chauvin, Christiaens, Currie, Danielski, Fortney, Girard, Greenbaum, Henning,
  Hines, Janson, Kalas, Kammerer, Kenworthy, Kervella, Lagage, Lew, Liu,
  Macintosh, Marino, Marley, Marois, Matthews, Matthews, Mawet, McElwain,
  Metchev, Meyer, Molliere, Pantin, Rebollido, Ren, Vasist, Wyatt, Zhou,
  Briesemeister, Bryan, Calissendorff, Catalloube, Cugno, De~Furio, Dupuy,
  Factor, Faherty, Fitzgerald, Franson, Gonzales, Hood, Howe, Kraus, Kuzuhara,
  Lawson, Lazzoni, Liu, Llop-Sayson, Lloyd, Martinez, Mazoyer, Quanz, Redai,
  Samland, Schlieder, Tamura, Tan, Uyama, Vigan, Vos, Wagner, Wolff, Ygouf,
  Zhang, \& Zhang}]{ref:22MiBiPa}
Miles, B.~E., Biller, B.~A., Patapis, P., {et~al.} 2022, The JWST Early Release
  Science Program for Direct Observations of Exoplanetary Systems II: A 1 to 20
  Micron Spectrum of the Planetary-Mass Companion VHS 1256-1257 b

\bibitem[{{Patapis} {et~al.}(2022){Patapis}, {Nasedkin}, {Cugno}, {Glauser},
  {Argyriou}, {Whiteford}, {Molli{\`e}re}, {Glasse}, \& {Quanz}}]{ref:22PaNaCu}
{Patapis}, P., {Nasedkin}, E., {Cugno}, G., {et~al.} 2022, \aap, 658, A72

\bibitem[{{Petrus} {et~al.}(2021){Petrus}, {Bonnefoy}, {Chauvin}, {Charnay},
  {Marleau}, {Gratton}, {Lagrange}, {Rameau}, {Mordasini}, {Nowak}, {Delorme},
  {Boccaletti}, {Carlotti}, {Houll{\'e}}, {Vigan}, {Allard}, {Desidera},
  {D'Orazi}, {Hoeijmakers}, {Wyttenbach}, \& {Lavie}}]{ref:21PeBoCh}
{Petrus}, S., {Bonnefoy}, M., {Chauvin}, G., {et~al.} 2021, \aap, 648, A59

\bibitem[{{Porto de Mello} {et~al.}(2014){Porto de Mello}, {da Silva}, {da
  Silva}, \& {de Nader}}]{ref:14MeSiSi}
{Porto de Mello}, G.~F., {da Silva}, R., {da Silva}, L., \& {de Nader}, R.~V.
  2014, \aap, 563, A52

\bibitem[{{Reach} {et~al.}(2005){Reach}, {Megeath}, {Cohen}, {Hora}, {Carey},
  {Surace}, {Willner}, {Barmby}, {Wilson}, {Glaccum}, {Lowrance}, {Marengo}, \&
  {Fazio}}]{ref:05ReMeCo}
{Reach}, W.~T., {Megeath}, S.~T., {Cohen}, M., {et~al.} 2005, \pasp, 117, 978

\bibitem[{Rigby {et~al.}(2022)Rigby, Perrin, McElwain, Kimble, Friedman, Lallo,
  Doyon, Feinberg, Ferruit, Glasse, {et~al.}}]{ref:22jwst}
Rigby, J., Perrin, M., McElwain, M., {et~al.} 2022, Characterization of JWST
  science performance from commissioning

\bibitem[{{Rothman} {et~al.}(2010){Rothman}, {Gordon}, {Barber}, {Dothe},
  {Gamache}, {Goldman}, {Perevalov}, {Tashkun}, \& {Tennyson}}]{ref:10RoGoBa}
{Rothman}, L.~S., {Gordon}, I.~E., {Barber}, R.~J., {et~al.} 2010, \jqsrt, 111,
  2139

\bibitem[{Sloan {et~al.}(2015)Sloan, Herter, Charmandaris, Sheth, Burgdorf, ,
  \& Houck}]{ref:15SlHeCh}
Sloan, G.~C., Herter, T.~L., Charmandaris, V., {et~al.} 2015, AJ, 149

\bibitem[{{Soubiran} \& {Triaud}(2004)}]{ref:04SoTr}
{Soubiran}, C. \& {Triaud}, A. 2004, \aap, 418, 1089

\bibitem[{Wells {et~al.}(2015)Wells, Pel, Glasse, Wright, Aitink-Kroes,
  Azzollini, Beard, Brandl, Gallie, Geers, Glauser, Hastings, Henning, Jager,
  Justtanont, Kruizinga, Lahuis, Lee, Martinez-Delgado, Martínez-Galarza,
  Meijers, Morrison, Müller, Nakos, O'Sullivan, Oudenhuysen, Parr-Burman,
  Pauwels, Rohloff, Schmalzl, Sykes, Thelen, van Dishoeck, Vandenbussche,
  Venema, Visser, Waters, \& Wright}]{ref:15WePeGl}
Wells, M., Pel, J.~W., Glasse, A., {et~al.} 2015, PASP, 127, 646

\bibitem[{{Wright} {et~al.}(2023){Wright}, {Rieke}, \& {Glasse}}]{ref:23WrRiGl}
{Wright}, G.~S., {Rieke}, G.~H., \& {Glasse}. 2023, \pasp, accepted

\bibitem[{{Yang} {et~al.}(2022){Yang}, {Green}, {Pontoppidan}, {Bergner},
  {Cleeves}, {Evans}, {Garrod}, {Jin}, {Kim}, {Kim}, {Lee}, {Sakai},
  {Shingledecker}, {Shope}, {Tobin}, \& {van Dishoeck}}]{ref:corinos}
{Yang}, Y.-L., {Green}, J.~D., {Pontoppidan}, K.~M., {et~al.} 2022, \apjl, 941,
  L13

\end{thebibliography}

\begin{appendix} 

\section{Comparison of periodograms}
\label{app:apppendixA}

\begin{figure*}[h!]
    \centering
    \includegraphics[width=\textwidth]{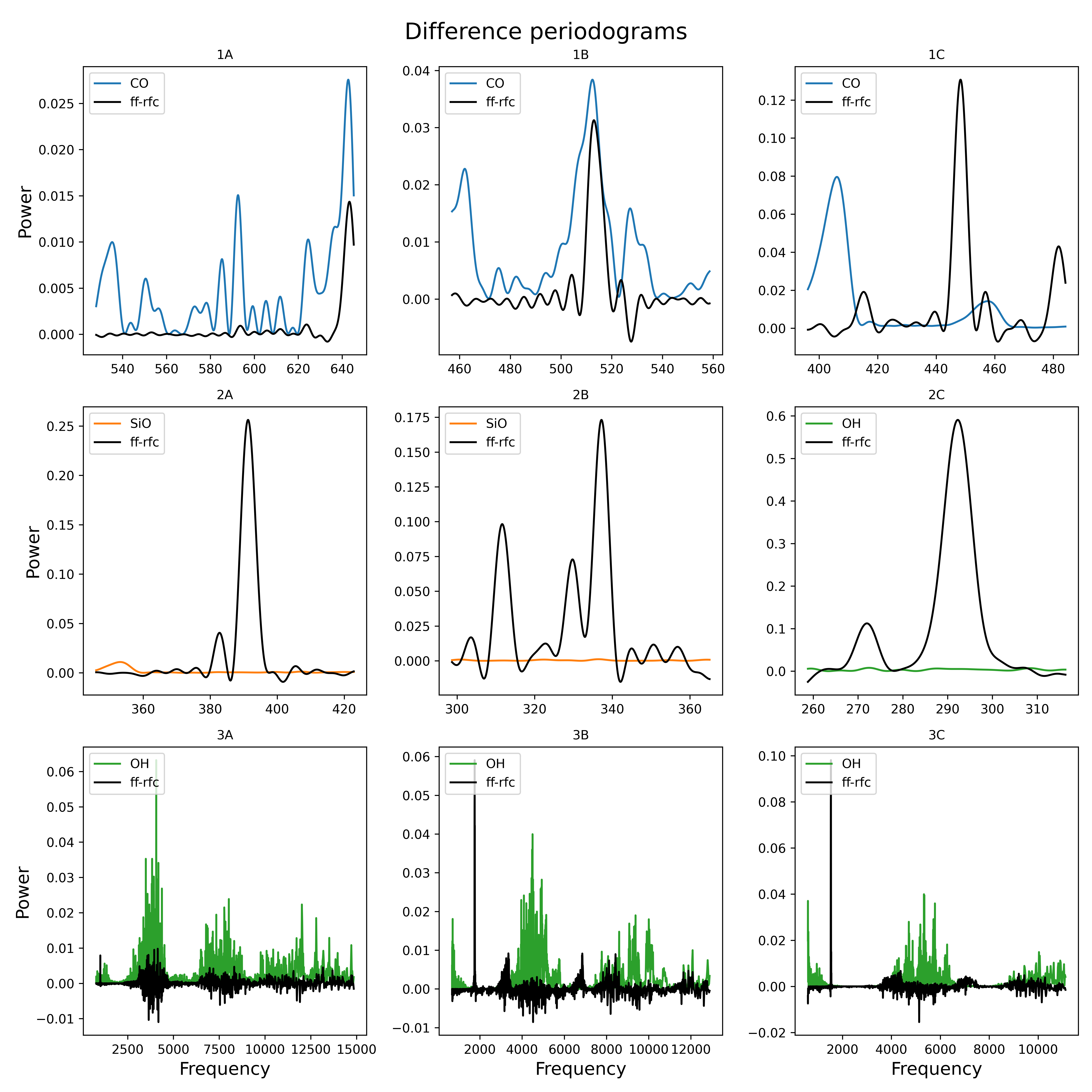}
    \caption{Difference between the periodograms of the K giant spectra before and after residual fringe correction per band, compared to the synthetic CO spectrum. The curves are colour-coded (see insets): `ff' stands for extended fringe flat and `rfc' for extended fringe flat+residual fringe correction. The section shown is the section where the algorithm searches for strong frequencies to remove.}
    \label{fig:diff_periodogram}
\end{figure*}

In this Appendix the additional periodograms of the K giant spectra and the molecular synthetic spectra are shown in Fig. \ref{fig:diff_periodogram}. Rather than presenting the periodograms of the spectrum including only the extended fringe flat and the extended fringe flat plus residual fringe correction separately, the difference between the two is represented by the black line. Each panel shows a different band, and the different coloured lines correspond to the periodograms of the molecules. Since the black line shows the difference between the two corrections, the influence of the residual fringe correction on the periodicity of the spectrum can be seen directly. The top row of the figure is analogous to Fig. \ref{fig:periods}. When the largest difference (i.e. the strongest peak) overlaps with a strong peak of the molecular periodogram, it is likely that the features were altered in the spectra.

This effect is seen in bands 1A and 1B, as was already mentioned prior in Sect. \ref{fig:contrast}, while this is not the case in 1C. In channel 2 none of the molecular periodograms show strong peaks in the altered frequencies, and are therefore likely not affected. In the longer wavelength channels, here channel 3, the search space is much wider, since a longer period beating on top of the shorter period fringe has to be corrected for. This makes the effect more difficult to see.

Additionally, it is especially evident in channel 3 that not only are frequencies removed, resulting in a positive difference, there are also frequencies added at a similar power to  those removed. This demonstrates that filtering out a sinusoid has more complex consequences for the spectrum than just removing a fringe.

\end{appendix}

\end{document}